\documentclass[12pt,aps,english]{revtex4}
\usepackage{graphicx}
\usepackage{bm}
\usepackage{dcolumn}
\usepackage[usenames, dvipsnames]{color}
\usepackage{epsfig}
\usepackage{amsmath}

\begin{document}

\title{Escape of a passive particle from activity-induced energy landscape: Emergence of slow and fast effective diffusion}
\author{Subhasish Chaki and Rajarshi Chakrabarti*}
\affiliation{Department of Chemistry, Indian Institute of Technology Bombay, Mumbai, Powai 400076, E-mail: rajarshi@chem.iitb.ac.in}

\begin{abstract}

\noindent Spontaneous persistent motions driven by active processes play a central role to maintain the living cells far from equilibrium. In the majority of the research works, the steady state dynamics of an active system has been described in terms of an effective temperature. By contrast, we have examined a prototype model for diffusion in an activity-induced rugged energy landscape to describe the slow dynamics of a tagged particle in a dense active environment. The expression for the mean escape time from the active rugged energy landscape holds only in the limit of low activity and the mean escape time from the rugged energy landscape increases with activity. The precise form of the active correlation will determine whether the mean escape time will depend on the persistence time or not. The active rugged energy landscape approach also allows an estimate of non-equilibrium effective diffusivity characterizing the slow diffusive motion of the tagged particle due to activity.  On the other hand, in a dilute environment,  high activity augments the diffusion of the tagged particle. The enhanced diffusion can be attributed to an effective temperature, higher than the ambient temperature and is used to calculate the Kramers' mean escape time, which decreases with activity. Our results have direct relevance to recent experiments on tagged particle diffusion in condensed phases.

\end{abstract}

\maketitle

\section{Introduction}

\noindent Systems composed of active particles represent a class of driven non-equilibrium systems in which the  driving forces are direct, isotropic and controlled locally rather than globally \cite{bechinger2016active, gnesotto2018broken}. They can either be found in biological systems such as  bacterial colonies \cite{berg2008coli}, motile cells in tissues \cite{gonzalez2012soft}, cytoskeleton in living cells \cite{ramaswamy2010mechanics} or are realized artificially such as catalytic Janus particles \cite{illien2017fuelled}. While the former are typically self-propelled by the chemical energy produced from the hydrolysis of adenosine triphosphate (ATP) \cite{brangwynne2008nonequilibrium}, the latter are coated with catalytic patches \cite{samin2015self} or illuminated by laser light \cite{buttinoni2012active} to achieve the directed motion. The activity-induced motion has been used in designing nano- and micron-sized machines \cite{ozin2005dream}, targeted drug delivery \cite{katuri2017designing} and active processes are believed to enhance chromatin mobility \cite{liu2018chain, jost2020}. However, a unified microscopic
 theoretical framework for active matter is still lacking. 
\\
\\
Breakdown of detailed balance or fluctuation-dissipation theorem, which makes the motion persistent, is a novel characteristic feature of active systems. That means the particle retains its direction of motion during a characteristic period of time. A somewhat simpler theoretical model, which  takes into account the persistence of active motion, is the Ornstein-Uhlenbeck process (OUP). In OUP, the self-propelled force varies randomly in both magnitude and direction. The OUP description of active matter has been applied in many theoretical studies \cite{ghosh2014dynamics, samanta2016chain, vandebroek2015dynamics, das2018confined, kumari2020stochastic, saha2019stochastic, feng2017mode} and provides accurate description of certain  experiments on living systems \cite{wu2000particle, maggi2014generalized}. In active bath, the dynamics of a tagged particle  becomes faster  due to athermal collisions from the active particles and leads to an enhanced diffusion in the long time limit \cite{argun2016non, samanta2016chain, chaki2019enhanced, 201szamel4self, chaki2019effects,aporvari2020anisotropic}. Hence the tagged particle in active bath is termed as ``Hot'' colloid \cite{chari2019scalar, argun2016non, bechinger2016active, gnesotto2018broken}. However, one can dig out an effective temperature \cite{chaki2019enhanced, bechinger2016active, 201szamel4self} from the enhanced diffusion using generalized fluctuation-dissipation theorem in active bath. In a simple manner, this effective temperature allows one to measure the energy scale of fluctuations coming from the active bath. The definition of effective temperature is not unique and it is different  depending on whether the active particle is confined or not \cite{201szamel4self, chaki2019effects}. The effective temperature description from an OUP is valid as long as the concentration of active particles is not so high as to give rise to collective phenomena \cite{bechinger2016active}. As a matter of fact, biological cells represent highly crowded  environment and hence activity plays a crucial role for structural rearrangement \cite{delarue2018mtorc1, parry2014bacterial} or efficient movement inside the cells \cite{bray2000cell, goychuk2014molecular}.  In order to understand the functioning of biological cells, recently the idea of effective temperature has also been extended to active glasses \cite{berthier2019glassy,  doi:10.1063/1.3624753} and higher the effective temperature, more ordered is the structure \cite{szamel2015glassy, flenner2016nonequilibrium}. In addition, there are cases where``active frozen states" arise from the competition between activity driven self-organization and growth \cite{bauschpnas2011, reichhardtpnas2011}. Such driven systems do not remain in a highly fluctuating state but rather self-organize into  long-lived disordered structures in which the fluctuations are strongly reduced. A good example of such steady state is a system of actin filaments that are actively transported by motor proteins and cross- linked in a two-dimensional geometry of a motility assay.  For dense soft glasses, the long time dynamics is always enhanced with increase in the effective temperature leading to the fluidization of the system \cite{mandal2016active}. Nandi $et. \, al.$ have recently extended the the random first-order transition (RFOT) theory to a dense assembly of self-propelled particles. In this case, the effects of activity have been accommodated in a renormalized potential energy density  that has a linear dependence on effective temperature \cite{nandi2018random}. In an enlightening paper, Leocmach et.al. have experimentally studied the response of a dense sediment of Brownian particles to self-propulsion \cite{klongvessa2019active}. In that experiment, the relaxation time unexpectedly increases in the very first non-zero activity and then decreases at high enough activity. In dense active systems, each particle is confined by its neighbors, leading to the formation of local potential barriers. Due to activity, the surroundings of a tagged particle undergo reorganization and hence the tagged particle is expected to escape from the local potential barriers of different heights. Epithelical cells also show a transition from fluid-like to glass-like regime due to maturation of cell-cell and cell-substrate contacts, as time progresses \cite{govpnas2015}. A few recent works have been devoted to the extension of the Kramers' escape \cite{kramers1940brownian} from a metastable state for active systems  \cite{sharma2017escape, woillez2019activated, geiseler2016kramers, caprini2019active,Woillezactivetrap}.   More recently, the notion of effective potential has emerged as an alternative approach to describe the Kramers' escape under self-propulsion rather than effective temperature \cite{wexler2020dynamics}. In another study, Pierro $et. \, al.$ have built a quasi-equilibrium energy landscape model  from optical experiments on chromatin \cite{di2018anomalous}. Though the energy landscape model does not have any activity dependent component, the temporal dynamics generated by this model significantly deviates from the equilibrium Rouse dynamics. Along the same line but in a different context, Volpe $et.\, al.$ have showed that fluctuation relations can not  be applied for a harmonically trapped colloidal particle in active bath \cite{argun2016non}. Interestingly,  they have restored these  fluctuation relations using an effective potential derived from the stationary distribution.

\begin{figure}[ht]
	\centering
	\begin{tabular}{cc}
		\includegraphics[width=1.10\textwidth]{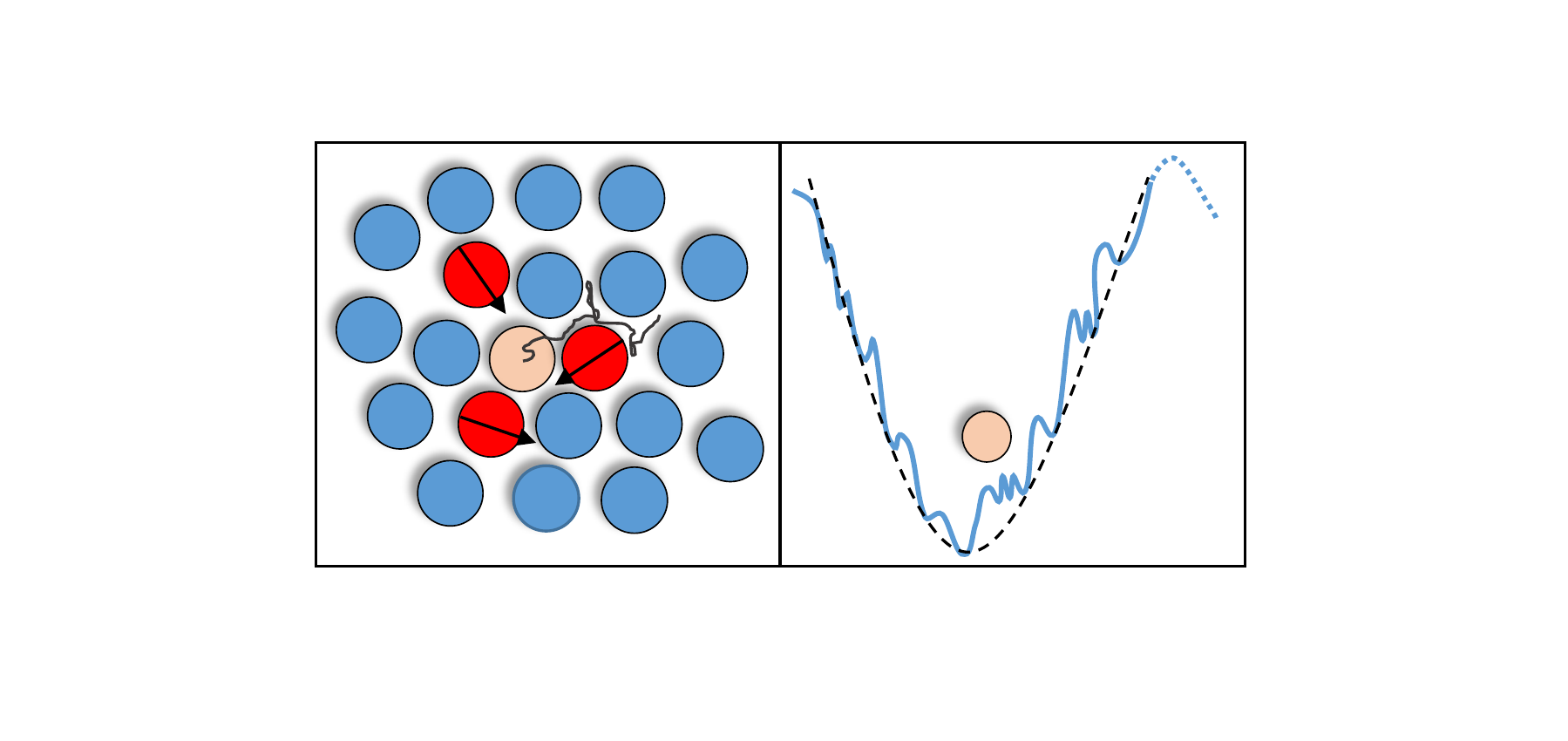} \\	
		                  (a)    		\\
			\includegraphics[width=1.1\textwidth]{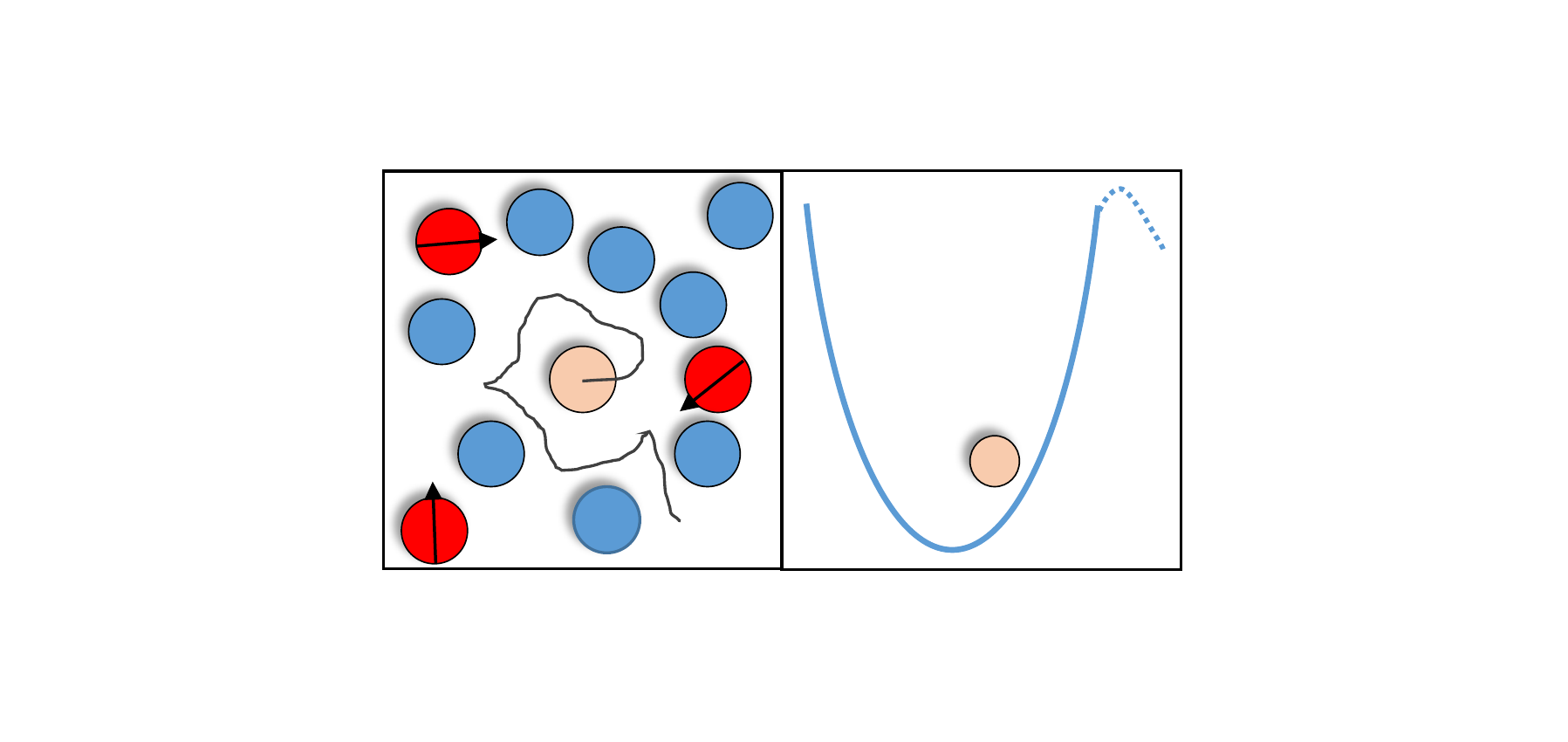} \\

                               (b)    \\				
	\end{tabular}
\caption{Schematic diagrams representing the AREL approach and effective temperature description. Dense mixture of active and passive particles are mapped to an AREL (Fig. (a)). Dilute mixture of active and passive particles are schematically described  in a smooth harmonic trap (Fig. (b)). The blue dotted lines in both the figures describe the inverted harmonic barrier tops.}
\label{fig:schematic}
\end{figure}

\clearpage
\newpage

\noindent Having discussed both the effective temperature and effective potential description in the context active matter, here we aim to investigate the escape of a tagged particle from a  trapping potential. The collection of interacting passive particles can be viewed as a single particle in an effective potential created by the neighboring particles. This effective potential will depend only on the single scalar dynamical variable, the radial displacement of the single particle in three dimensional space \cite{nandi2017role,schweizer2005derivation,schweizer2003entropic,mirigian2014elastically}. In a recent work \cite{nandi2017role}, it has been proposed that the mapping of interacting system of particles into a single particle in an effective field can be related to the microscopic mode coupling theory (MCT).   In our model, this effective potential is taken to be a harmonic well, which accounts for the trapping of the tagged particle due to surrounding particles and then an inverted harmonic barrier top at larger displacement of the tagged particle, which represents the saddle point and is eventually used to calculate the mean escape time \cite{kramers1990physica, chakrabarti2007exact, schweizer2005derivation,schweizer2003entropic}. We assume that a small fraction of particles is active and the net effect of activity  is introduced in the model by  an additional random force drawn from the steady state distribution of  OUP.  Thus a passive particle in such an environment will experience the usual trapping potential due to interaction with the neighboring particles, and an additional ruggedness in the potential arises due to active particles in the environment. Such disordered states can be termed as ``active frozen steady state" \cite{bauschpnas2011, reichhardtpnas2011}. To account for the dynamics of a tagged particle in this active frozen state, we make an extension of the Zwanzig's celebrated theory for rugged energy landscape to active systems \cite{zwanzig1988diffusion} and termed as active rugged energy landscape (AREL). We obtain an expression for mean first passage time (MFPT) in AREL and show that the MFPT is sensitive to the microscopic details of how the activity is implemented. To describe the slow dynamics of the tagged particle in dense environment, we define an effective diffusivity in AREL framework which decreases with increasing activity.  This is contrary to the earlier works where diffusion is always enhanced due to activity.  Importantly, the limitations of the AREL approach and the conditions for its validity are discussed in detail. Furthermore, by adopting the notion of effective temperature derived from the generalized fluctuation-dissipation theorem in active bath, we analyze the MFPT and provide a comparative discussion between AREL approach and effective temperature method. We schematically illustrate  the AREL framework and the effective temperature description in Fig. (\ref{fig:schematic}).

\section{Theoretical framework for AREL}

\noindent We consider a passive particle in a trap described by a potential $U(x)$ where $x$ is the position of the passive particle. The $U(x)$ is composed of a smooth background trapping potential $U_0(x)$ and a random potential $U_A(x)$ superimposed on $U_0(x)$.   The smooth potential $U_0(x)$ mimics the cytoskeleton confinement inside the biological cells \cite{gov2003cytoskeleton,li2007cytoskeletal,kumar2019transport} or in a different scenario, accounts for the averaged many-body potential energy coming from the interactions between surrounding particles \cite{shen2004stability,nandi2018random,samanta2016tracer}. The microscopic origin of $U_A(x)$ is attributed to the various complex active processes inside the system. In our model, we define $U_A(x) =x f_A $ where $f_A$ is the active force. Hence, $U(x)=U_0(x)+x f_A$ where the rugged part of the potential comes from the activity. In steady state, the values of $f_A$ are chosen from a time independent distribution. This situation refers to the scenario of ``static disorder'', as termed by Zwanzig  \cite{zwanzig1990rate}. In fact, $f_A$ is independent of the state ($x$) of the passive tagged particle.
\\
\\
In the high friction limit, the diffusive motion of the particle subjected to the trapping potential $U(x)$ is best described by the Smoluchowski equation,

\begin{equation}
\begin{split}
 \frac{\partial P(x,t|x_0,0)}{\partial t}&=-\frac{\partial J (x,t|x_0,0)}{\partial x}
\label{eq:active_fokker}
\end{split}
\end{equation} 

\begin{equation}
\begin{split}
J  (x,t|x_0,0)=-D \exp \left(-\beta U(x)\right) \frac{\partial }{\partial x} \left[\exp \left(\beta U(x)\right) P(x,t|x_0,0)\right]
\label{eq:active_flux}
\end{split}
\end{equation} 

\noindent where  $P(x,t|x_0,0)$ be the probability to find the system at point $x$ at time $t$, provided that it has started at $x_0$ at $t=0$,  $J$ is the flux and $\beta=\frac{1}{k_B T}$ where $k_B$ is the Boltzmann constant, and $T$ is the ambient temperature. The initial condition at $t=0$ is given by  $P(x,0|x_0,0)=\delta (x-x_0)$. The thermal diffusion constant, $D$  is related to friction coefficient $\gamma$ through fluctuation-dissipation relation, $D=\frac{k_B T}{\gamma}$.
\\
\\
The mean time required for the particle starting out at $x_0$ to reach $b$ for the first time is termed as mean first passage time (MFPT) and denoted by $\left<\tau (x_0)\right>$. Hence, the absorbing boundary is located at $x=b$ and consequently $P(x,t\rightarrow \infty|x_0,0)=0$. By the consideration of the boundary conditions imposed on $P(x,t|x_0,0)$ and solving the differential equation Eq. (\ref{eq:active_fokker}, \ref{eq:active_flux}), the following adjoint equation for $\tau(x_0)$ can be obtained: \cite{zwanzig2001nonequilibrium}

\begin{equation}
\begin{split}
D \exp \left(\beta U(x_0)\right) \frac{\partial }{\partial x_0} \left[ \exp \left(-\beta U(x_0)\right) \frac{\partial }{\partial x_0} \left<\tau (x_0)\right>\right]=-1
\label{eq:active_MFPT}
\end{split}
\end{equation} 

\noindent Solving the differential Eq. (\ref{eq:active_MFPT}) using $\left<\tau (b)\right>=0$, we arrive at the expression of MFPT, $\left<\tau (x_0)\right>$:

\begin{equation}
\begin{split}
 \left<\tau (x_0)\right>=\int_{x_0}^b dy \exp \left(\beta U(y)\right) \frac{1}{D} \int_{-\infty}^y dz \exp \left(-\beta U(z)\right)
\label{eq:active_tau}
\end{split}
\end{equation} 

\noindent Following the pioneering work by Zwanzig \cite{zwanzig1988diffusion}, the integral over a small distance can be approximated by,

\begin{equation}
\begin{split}
 \int dz \exp \left(-\beta U(z)\right) & \approx  \int dz \exp \left(-\beta U_0(z)\right) \left< \exp \left(-\beta U_A(z)\right)\right> \\
&= \int dz \exp \left(-\beta U_0(z)\right) \left< \exp \left(-\beta z f_A\right)\right>
\label{eq:active_tau_approx}
\end{split}
\end{equation} 

\noindent Here $\left<......\right>$ denotes the average over the steady state ensemble (or snapshots) of different local rearrangements in the environment. In our model, this formalism proposed by Zwanzig in ref. \cite{zwanzig1988diffusion} works well because the length scale of structural ruggedness in the landscape due to activity is much smaller than the length scale of diffusion. Hence, the average of the exponential will be a function of the coordinate $z$ and described as, 

\begin{equation}
\begin{split}
 \left< \exp \left(-\beta z f_A\right)\right>& = \int_{-\infty}^{\infty}df_A P(f_A) \exp \left(-\beta z f_A\right)  \\
& = \exp (\lambda_- (z))
\label{eq:active_tau_approx---}
\end{split}
\end{equation}

\noindent  This averaging is very similar in  spirit of  ``superstatistics"  as introduced by  Cohen and Beck \cite{beck2003superstatistics}. However, unlike the distribution of temperature as in case of superstatistics, here we have a distribution of ruggedness due to activity. In a similar way, one can approximate the other integral of the exponential function in Eq. (\ref{eq:active_tau_approx}), 

\begin{equation}
\begin{split}
 \left< \exp \left(\beta y f_A\right)\right> = \exp (\lambda_+ (y))
\label{eq:active_tau_approx+++}
\end{split}
\end{equation} 

\noindent Using Eq. (\ref{eq:active_tau_approx}, \ref{eq:active_tau_approx---}, \ref{eq:active_tau_approx+++}),  Eq. (\ref{eq:active_tau}) can be transformed into, 

\begin{equation}
\begin{split}
 \left<\tau (x_0)\right>=\int_{x_0}^b dy \exp \left(\beta U_0(y)+\lambda_+(y)\right) \frac{1}{D} \int_{-\infty}^y dz \exp \left(-\beta U_0(z)+\lambda_- (z)\right)
\label{eq:active_tau_final}
\end{split}
\end{equation}

\noindent This is the general expression of the MFPT for the rugged energy landscape.  To determine the MFPT, $\tau(x_0)$ in the presence of activity, we have to explicitly calculate  the steady state distribution of the active noise, $f_A$ that enters in Eq. (\ref{eq:active_tau_final}) through $\lambda_+$ and $\lambda_-$. 

\section{Different models of activity}

\noindent We consider two different active noise statistics \cite{nandi2017nonequilibrium},
\\
\\
\noindent \textbf{Model 1}: The fluctuations of the active noise $f_A(t)$ is governed by an Ornstein-Uhlenbeck (OU) process \cite{201szamel4self}, such that

\begin{equation}
\tau_A\dot{f_A} = -f_A(t)+  \Gamma(t)
\label{eq:ornstein}
\end{equation}

\noindent where $\tau_A$ is the persistence time of the active noise, $\Gamma(t)$ is a Gaussian noise with $\left<\Gamma(t)\right>=0$ and $\left\langle \Gamma(t) \Gamma(t^{\prime}) \right\rangle=2C_0 (T)\delta(t-t^\prime)$. Here, $C_0(T)$ can be any arbitrary function of the ambient temperature $T$  that breaks the detailed balance condition \cite{sandford2017pressure,nandi2018random}. This noise statistic applies to the motion of the colloidal particle in a bacterial bath, where both the direction and amplitude of the active force change gradually due to various interactions with the motile bacteria \cite{wu2000particle}.

\noindent From Eq. (\ref{eq:ornstein}), one can show that 

\begin{equation}
\left<f_A(t)\right> = f_A(0) \exp\left[-\frac{t}{\tau_A}\right]
\label{eq:ornstein_mean}
\end{equation}

\noindent where $f_A(0)$ is the initial value of the active force. From Eq. (\ref{eq:ornstein_mean}), it is evident that the system reaches the steady state with the time scale as characterized by $\tau_A$. In the steady state it can be shown that 

\begin{equation}
\left<f_A(t) f_A(t^\prime)\right> = \frac{C_0}{\tau_A} \exp\left[-\frac{|t-t^\prime|}{\tau_A}\right]
\label{eq:ornstein_corr}
\end{equation}

\noindent In the limit $\tau_A\rightarrow 0$, the active noise has no memory and is $\delta$-correlated in time. In other words, it is equivalent to re-scaling the ambient temperature. Hence, the steady state distribution of the active noise will be

\begin{equation}
P_s(f_A) = N  \exp\left(-\frac{\tau_A f_A^2}{C_0}\right)
\label{eq:ornstein_distribution}
\end{equation}

\noindent where the normalization constant, $N$ is $\left(\frac{\tau_A}{C_0 \pi}\right)^{\frac{1}{2}}$.

\noindent \textbf{Model 2}: Here the active force, $f_A$ has telegraphic-noise temporal correlation, 

\begin{equation}
\left<f_A(t) f_A(t^\prime)\right> = f_B^2 \exp\left[-\frac{|t-t^\prime|}{\tau_B}\right]
\label{eq:ornstein_corr_shot}
\end{equation}

\noindent This realization of the active noise is considered to study the active processes inside the cytoplasm of a cell, where activity arises from the interactions of the molecular motors \cite{ben2011effective}. In this model, the active particles produce pulses of average force $f_B$ for a constant duration $\tau_B$. For telegraphic noise driven processes, the pulses are turned on randomly as a Poisson process with an average waiting time $\tau$. Hence, $P(f_A)$ will be non-Gaussian \cite{goswami2019diffusion, um2019langevin, chaki2019enhanced}. However, for small $\tau$ and $\tau_B$, the time evolution of $f_A$ can be approximately mapped to an OUP, 

\begin{equation}
\dot{f_A} = - \frac{f_A(t)}{\tau_B}+ \frac{ \Gamma(t)}{\sqrt{\tau_B}}
\label{eq:ornstein_shot}
\end{equation}

\noindent where $\Gamma(t)$ is a standard Gaussian noise with statistical properties, $\left<\Gamma(t)\right>=0$ and $\left\langle \Gamma(t) \Gamma(t^{\prime}) \right\rangle=2f_B^2\delta(t-t^\prime)$. For this case in the steady state,  $P(f_A)$ will be $P_s(f_A)$, such that 

\begin{equation}
P_s(f_A) = N  \exp\left(-\frac{ f_A^2}{f_B^2}\right)
\label{eq:ornstein_distribution_shot}
\end{equation}

\noindent where the normalization constant, $N$ is $\frac{1}{f_B\sqrt{\pi}}$.  For $\tau_B \rightarrow 0$, the active noise correlation vanishes in Eq. (\ref{eq:ornstein_corr_shot}). Hence  Model 2 differs from  Model 1.
\\
\\
One should note that $P_s(f_A)$ for both the models is Gaussian  even though the system is far from equilibrium \cite{chaki2018}. For Gaussian distribution of $f_A$, both  $\exp (\lambda_- (z))$ and $\exp (\lambda_+ (y))$ will have the same expression and the sign of $U_A(x)$ will be a matter of convention. For model 1, $\exp (\lambda_- (z))=\exp \left(\frac{\beta^2 z^2 C_0}{4\tau_A}\right)$ and for model 2, $\exp (\lambda_- (z))=\exp \left(\frac{\beta^2 z^2 f_B^2}{4}\right)$.  In  model 1 , the persistence time, $\tau_A$  is independent of the the activity, $C_0$. The same is true for  model 2.

\section{Barrier crossing in AREL}

\noindent In this section, we will explore Eq. (\ref{eq:active_tau_final}) in the presence of activity for confining potentials. When $k_B T$ is small, the predominant contribution to the integration over $z$ in Eq.  (\ref{eq:active_tau_final}) comes from the immediate neighborhood of $a$ where $a$ is a simple minimum of $U_0(z)$,  

\begin{equation}
U_0(z)=U_{\textrm{min}}+\frac{1}{2}m \omega_{\textrm{min}}^2 \left(z-a\right)^2 +......
\label{eq:potential_z}
\end{equation}

\noindent where $U_{\textrm{min}}=\frac{1}{2}m \omega_{\textrm{min}}^2 a^2$. Next we extend the upper limit of the integration from $y$ to $\infty$ and the integral for model 1 is, 

\begin{equation}
\begin{split}
& \int_{-\infty}^{\infty} dz \exp \left(-\beta\left( U_{\textrm{min}}+\frac{1}{2}m \omega_{\textrm{min}}^2 \left(z-a\right)^2\right)+\frac{\beta^2 z^2 C_0}{4\tau_A}\right)  \\
&=\exp \left(-2\beta U_{\textrm{min}}+\frac{\beta^2 U_{\textrm{min}}^2}{a^2 \Omega_1}\right)  \sqrt{\frac{\pi}{\Omega_1}} \\ 
\label{eq:potential_z_integration}
\end{split}
\end{equation}

\noindent where $\Omega_1=\left(\frac{1}{2}m \beta \omega_{\textrm{min}}^2-\frac{\beta^2  C_0}{4\tau_A}\right)$. The absorbing barrier is placed at the maximum, $x=b$, of the potential $U(x)$. The integral over $y$ is dominated by the potential near the absorbing barrier and we will follow the same procedure previously done for the integral over $z$.  Hence, the expansion will be quadratic, 

\begin{equation}
U_0(y)=U_{\textrm{max}}-\frac{1}{2}m \omega_{\textrm{max}}^2 \left(y-b\right)^2 
\label{eq:potential_y}
\end{equation}

\noindent where $U_{\textrm{max}}=\frac{1}{2}m \omega_{\textrm{max}}^2 b^2$ and set the range of integration from $-\infty$ to $\infty$. Thus, 

\begin{equation}
\begin{split}
& \int_{-\infty}^{\infty} dy \exp \left(\beta\left( U_{\textrm{max}}-\frac{1}{2}m \omega_{\textrm{max}}^2 \left(y-b\right)^2\right)+\frac{\beta^2 y^2 C_0}{4\tau_A}\right)  \\
&=\exp \left(\frac{\beta^2 U_{\textrm{max}}^2}{b^2 \Omega_2}\right)  \sqrt{\frac{\pi}{\Omega_2}} \\
\label{eq:potential_y_integration}
\end{split}
\end{equation}

\noindent where $\Omega_2=\left(\frac{1}{2}m \beta \omega_{\textrm{max}}^2-\frac{\beta^2  C_0}{4\tau_A}\right)$. Substituting Eq. (\ref{eq:potential_z_integration}, \ref{eq:potential_y_integration}) into Eq. (\ref{eq:active_tau_final}) and replacing $\left<\tau (x_0)\right>$ by $\left<\tau^1_U\right>$, we obtain

\begin{equation}
\begin{split}
 \left<\tau^1_U\right>&=\frac{\pi}{D\sqrt{\Omega_1 \Omega_2}} \exp \left(-2\beta U_{\textrm{min}}+\frac{\beta^2 U_{\textrm{min}}^2}{a^2 \Omega_1}+\frac{\beta^2 U_{\textrm{max}}^2}{b^2 \Omega_2}\right)
\label{eq:active_tau_model_1}
\end{split}
\end{equation}

\begin{figure*}[ht]
\begin{center}
\begin{tabular}{cc}
\includegraphics[width=0.55\textwidth]{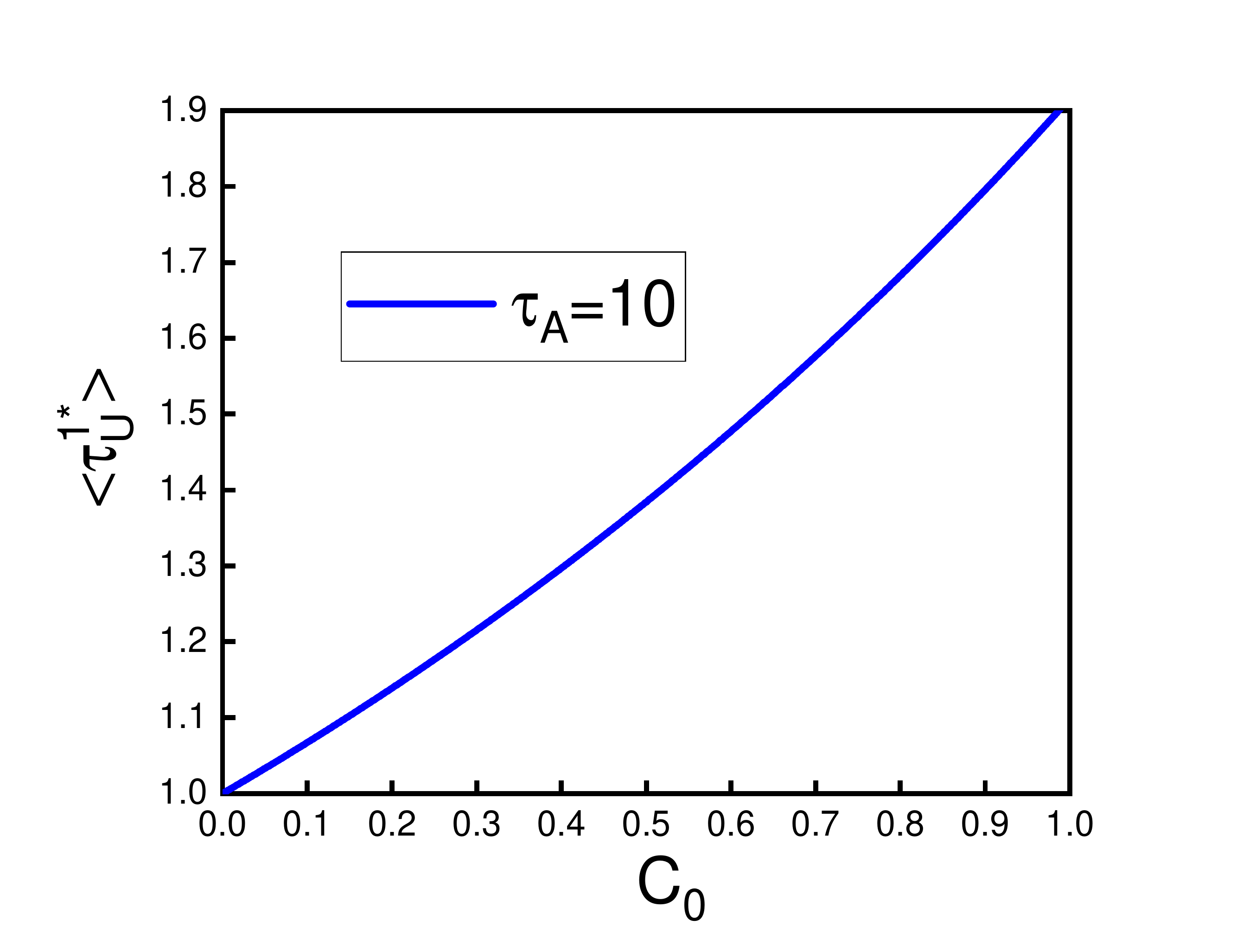} &
\includegraphics[width=0.55\textwidth]{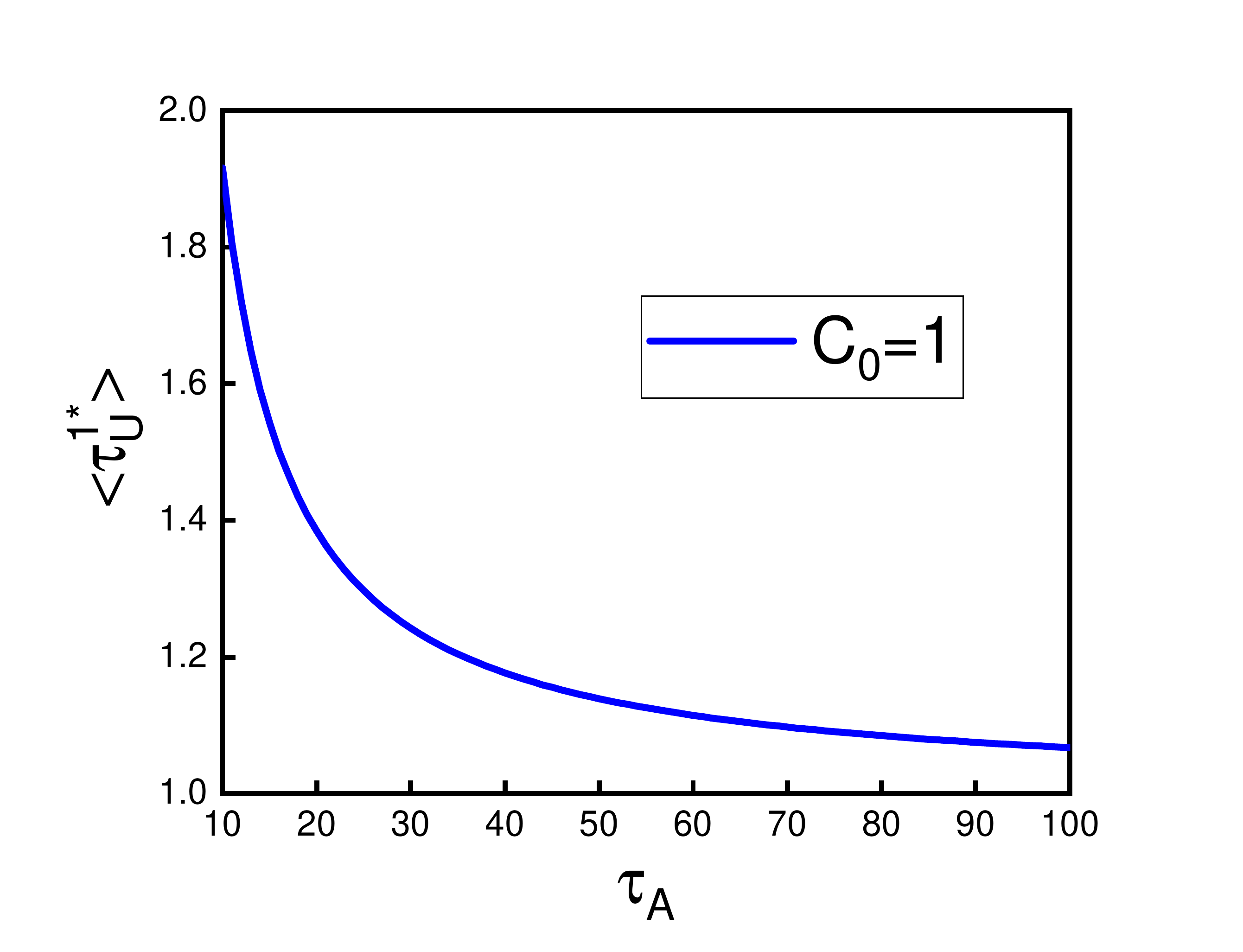}\\
(a) & (b)  \\
\end{tabular}
\end{center}
\caption{Plots of $\left<\tau^{1*}_U \right>$ against $C_0$ (plot (a)) and $\tau_A$ (plot (b)) for single particle in AREL. The values of the parameters used for both the plots are   $m=1, \gamma=1, k_B=1, T=1, a=1, b=5, \omega_{\textrm{min}}=10, \omega_{\textrm{max}}=20$.}
\label{fig:active_energy_A}
\end{figure*}

\noindent In a similar way, one can calculate the MFPT, $ \left<\tau^2_U\right>$ for model 2, 

\begin{equation}
\begin{split}
 \left<\tau^2_U\right>&=\frac{\pi}{D\sqrt{\xi_1 \xi_2}} \exp \left(-2\beta U_{\textrm{min}}+\frac{\beta^2 U_{\textrm{min}}^2}{a^2 \xi_1}+\frac{\beta^2 U_{\textrm{max}}^2}{b^2 \xi_2}\right)
\label{eq:active_tau_model_2}
\end{split}
\end{equation} 

\noindent where $\xi_1=\left(\frac{1}{2}m \beta \omega_{\textrm{min}}^2-\frac{\beta^2  f_B^2}{4}\right)$ and $\xi_2=\left(\frac{1}{2}m \beta \omega_{\textrm{max}}^2-\frac{\beta^2  f_B^2}{4}\right)$.

\begin{figure*}[h]
\begin{center}
\begin{tabular}{cc}
\includegraphics[width=0.55\textwidth]{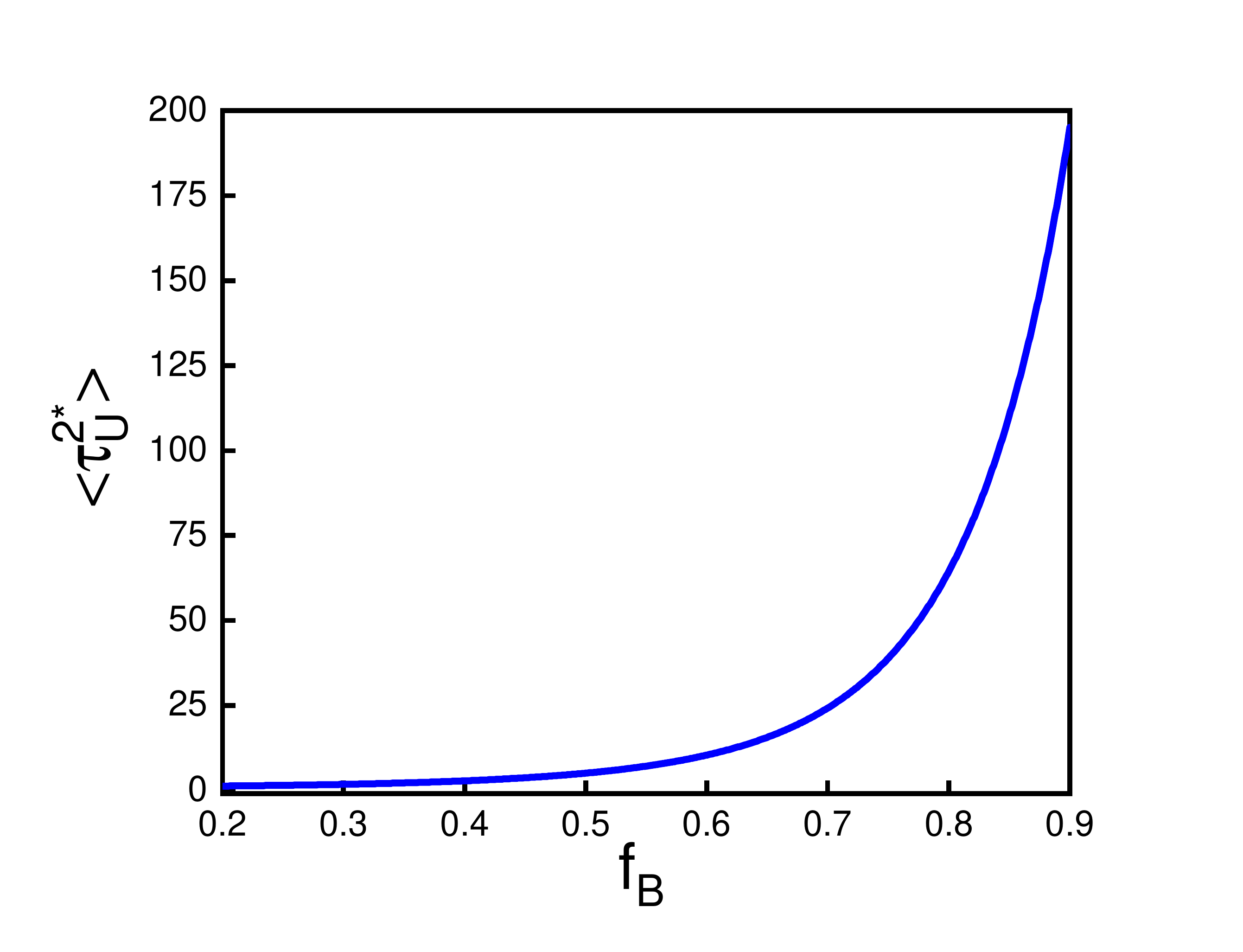} 
\end{tabular}
\end{center}
\caption{Plot of $\left<\tau^{2*}_U\right>$ against $f_B$  for single particle in AREL. The values of the parameters used for the plot are   $m=1, \gamma=1, k_B=1, T=1, a=1, b=5, \omega_{\textrm{min}}=10, \omega_{\textrm{max}}=20$.}
\label{fig:active_energy_B}
\end{figure*}

\noindent For both the models 1 and 2, the modified MFPTs $\left<\tau^1_U \right>$ and $\left<\tau^2_U \right>$ in the presence of activity, significantly deviate from the Kramer's theory based on thermal equilibrium conditions. Variation of non-dimensional MFPT $\left(\left<\tau^{1*}_U \right>=\frac{\left<\tau^1_U \right>}{\left<\tau_T \right>}\right)$ against $C_0$ and $\tau_A$ can be seen in Fig. \ref{fig:active_energy_A} where $\left<\tau_T \right>$ is the Kramers' MFPT. The AREL framework for model 1 shows that $\left<\tau^{1*}_U \right>$ increases with increasing the activity $C_0$ for constant $\tau_A$ ( Fig. \ref{fig:active_energy_A} (a)). Here activity slows down the dynamics which seems counter-intuitive and emerges from the fact that activity brings ruggedness. On the other hand at fixed $C_0$, $\left<\tau^{1*}_U \right>$ decreases monotonically with $\tau_A$ ( Fig. \ref{fig:active_energy_A} (b)). Hence, larger $\tau_A$ drives the trapped particle away from the potential minimum, $a$  for longer times.  With increasing $f_B$, $\left<\tau^{2*}_U \right>$ grows for model 2 (Fig. \ref{fig:active_energy_B}) but at a slower rate as compared to Fig. \ref{fig:active_energy_A} (a). However the MFPT (Eq. (\ref{eq:active_tau_model_2})) is independent of the persistence time $\tau_B$. This is perceived as contradictory with the work by Caprini $et. al.$ \cite{caprini2019active} where persistence time play a crucial role for barrier crossing.  Hence, the effect of decreasing $\tau_A$ on ruggedness in model 1, ought to be compared against increase in $C_0$ and $f_B$.   It should be noted that the prefactors of the exponentials in Eq. (\ref{eq:active_tau_model_1}, \ref{eq:active_tau_model_2}) have an unusual dependence on the ambient temperature, $T$ whereas the  prefactor of the exponential in Kramers' theory is independent of $T$.

\begin{figure*}[ht]
\begin{center}
\begin{tabular}{cc}
\includegraphics[width=0.55\textwidth]{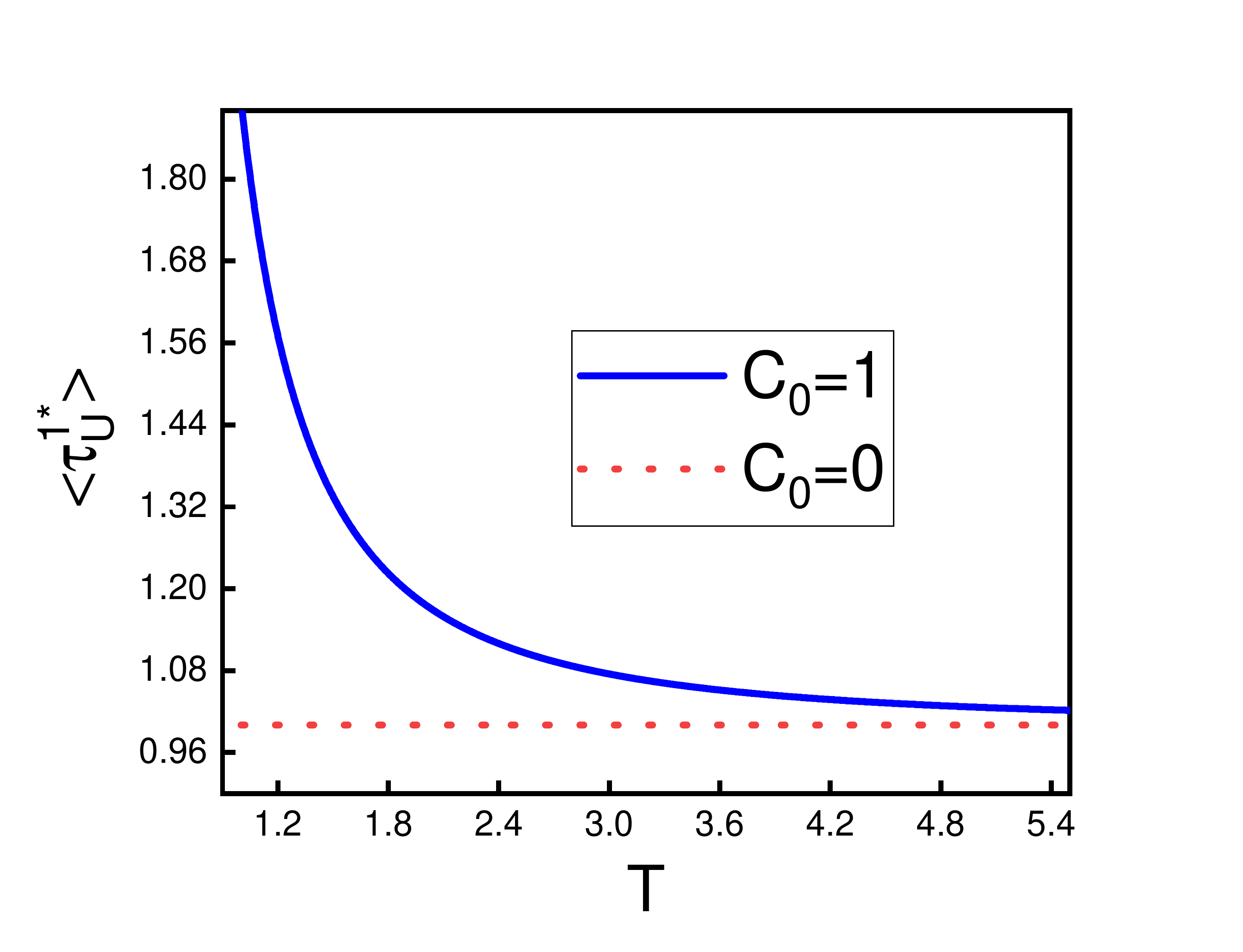} &
\includegraphics[width=0.55\textwidth]{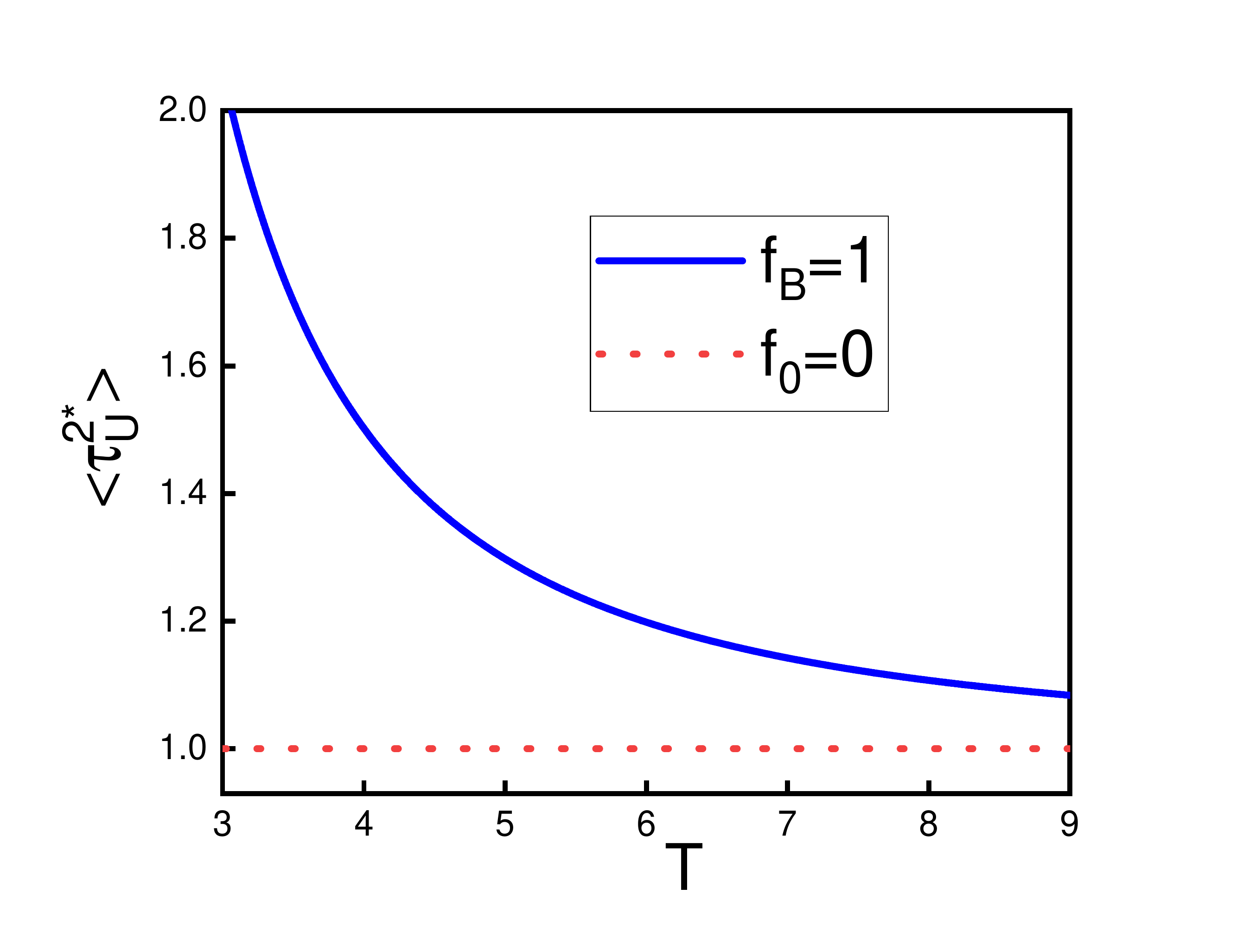}\\
(a) & (b)  \\
\end{tabular}
\end{center}
\caption{Plots of $\left<\tau^{1*}_U \right>$ and $\left<\tau^{2*}_U \right>$ against $T$ with different activity for single particle in AREL for models 1 (plot (a)) and 2 (plot (b)). The values of the parameters used for both the plots are   $m=1, \gamma=1, k_B=1, a=1, b=5, \omega_{\textrm{min}}=10, \omega_{\textrm{max}}=20, \tau_A=10$.} 
\label{fig:active_energy_T}
\end{figure*}

\noindent In Fig. (\ref{fig:active_energy_T}), the evolving natures of $\left<\tau^{1*}_U \right>$ and $\left<\tau^{2*}_U \right>$ with temperature $T$  are quite different. However, for small values of $T$, the MFPT in the presence of activity deviates from the equilibrium Kramers' MFPT (red dotted lines). At higher $T$ or small $\beta$, we can approximately  write $\Omega_2=\left(\frac{1}{2}m \beta \omega_{\textrm{max}}^2-\frac{\beta^2  C_0}{4\tau_A}\right) \approx \frac{1}{2}m \beta \omega_{\textrm{max}}^2$ and $\Omega_1=\left(\frac{1}{2}m \beta \omega_{\textrm{min}}^2-\frac{\beta^2  C_0}{4\tau_A}\right) \approx \frac{1}{2}m \beta \omega_{\textrm{min}}^2$ for model 1. Similarly for model 2 at higher $T$, $\xi_1 \approx \frac{1}{2}m \beta \omega_{\textrm{min}}^2$ and $\xi_2 \approx \frac{1}{2}m \beta \omega_{\textrm{max}}^2$. Hence, the dynamics is dominated by thermal noise for both models 1 and 2 and the probability of getting trapped in the metastable states will be small. As a result, the active MFPT curve asymptotically merges with the equilibrium MFPT result (red dotted line) in Fig. (\ref{fig:active_energy_T}).
\\
\\
\noindent In Eq. (\ref{eq:active_tau_model_1}, \ref{eq:active_tau_model_2}), the MFPTs $\left<\tau^{1}_U \right>$ and $\left<\tau^{2}_U \right>$ are exponential in nature due to the Gaussian distribution of active noise for both the models 1 and 2. This allows us to describe the system with an effective potential energy for both the models 1 and 2 but keeping the same pre-factor  that of equilibrium Kramers' result, 

\begin{equation}
\begin{split}
 \left<\tau^1_U \right>&=\frac{2 \pi \gamma}{m \omega_{\textrm{min}} \omega_{\textrm{max}}} \exp \left(\beta \Delta U_{\textrm{eff}}^1\right)
\label{eq:active_tau_model_1U}
\end{split}
\end{equation} 

\begin{equation}
\begin{split}
 \left<\tau^2_U \right>&=\frac{2 \pi \gamma}{m \omega_{\textrm{min}} \omega_{\textrm{max}}} \exp \left(\beta \Delta U_{\textrm{eff}}^2\right)
\label{eq:active_tau_model_2U}
\end{split}
\end{equation} 

\begin{figure*}[ht]
\begin{center}
\begin{tabular}{cc}
\includegraphics[width=0.55\textwidth]{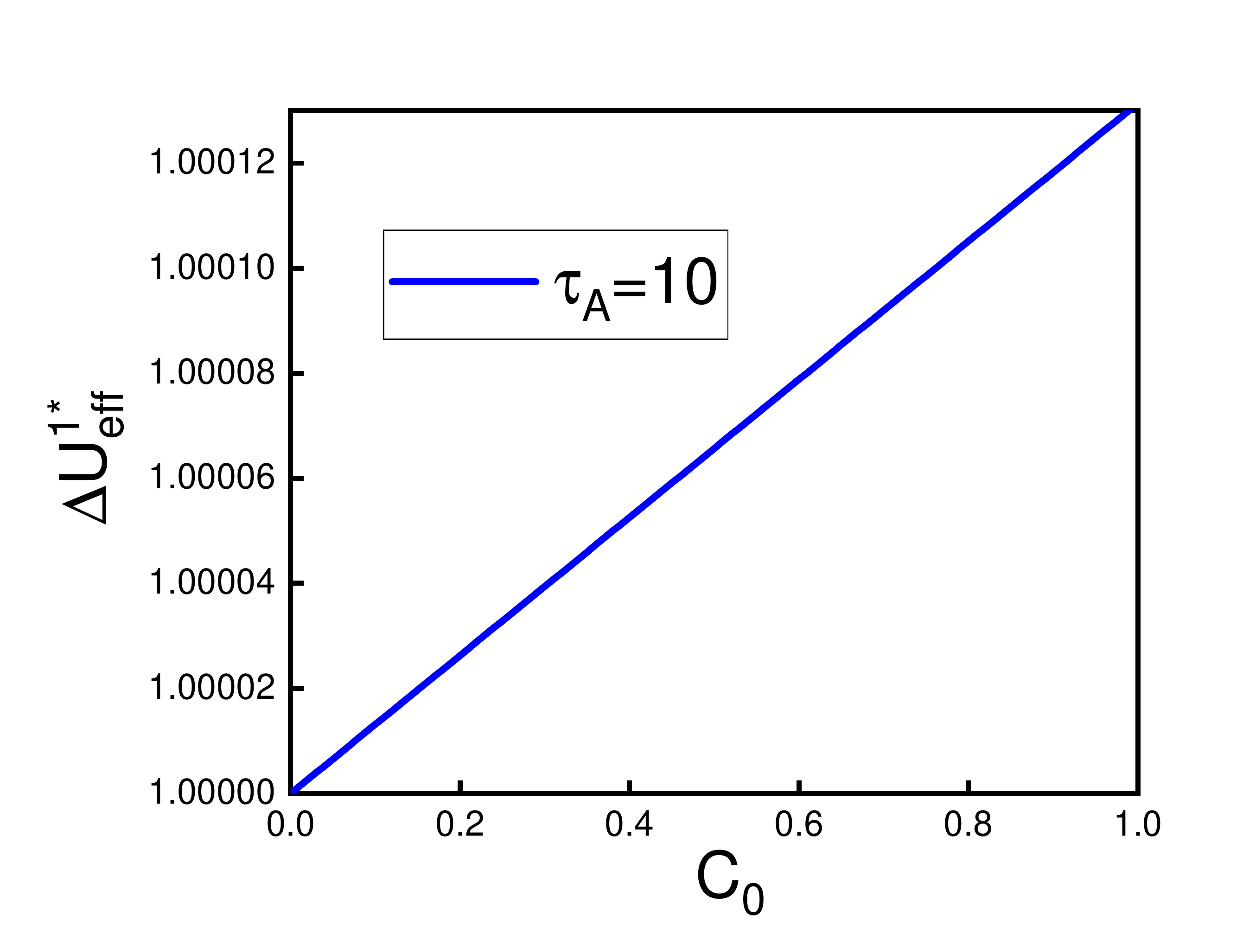} &
\includegraphics[width=0.55\textwidth]{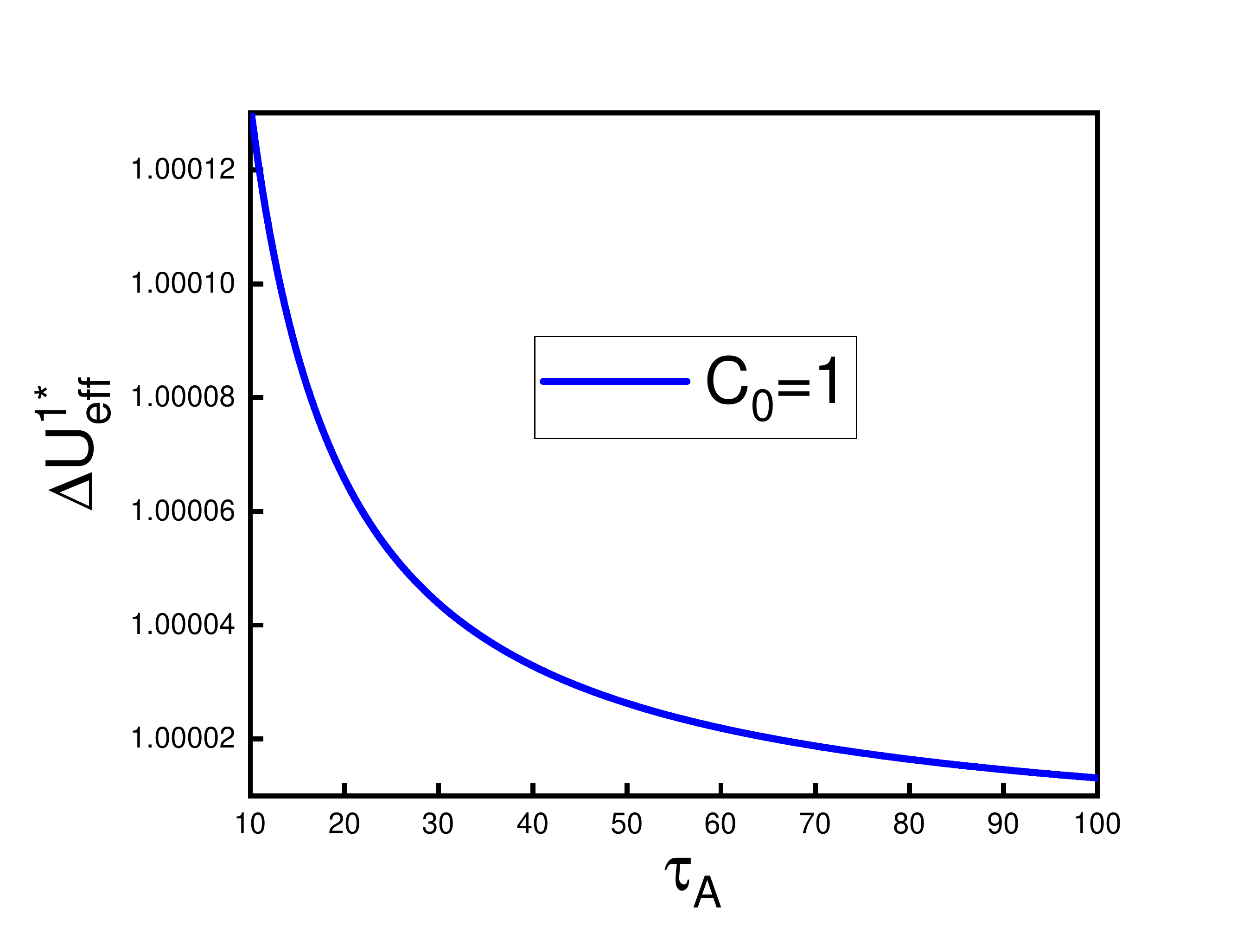}\\
(a) & (b)  \\
\end{tabular}
\end{center}
\caption{Plots of $\Delta  U_{\textrm{eff}}^{1*}$ against $C_0$ (plot (a)) and $\tau_A$ (plot (b)) for single particle in AREL. The values of the parameters used for both the plots are   $m=1, \gamma=1, k_B=1, T=1, a=1, b=5, \omega_{\textrm{min}}=10, \omega_{\textrm{max}}=20$.}
\label{fig:active_potential_height_A}
\end{figure*}

\begin{figure*}[h]
\begin{center}
\begin{tabular}{cc}
\includegraphics[width=0.55\textwidth]{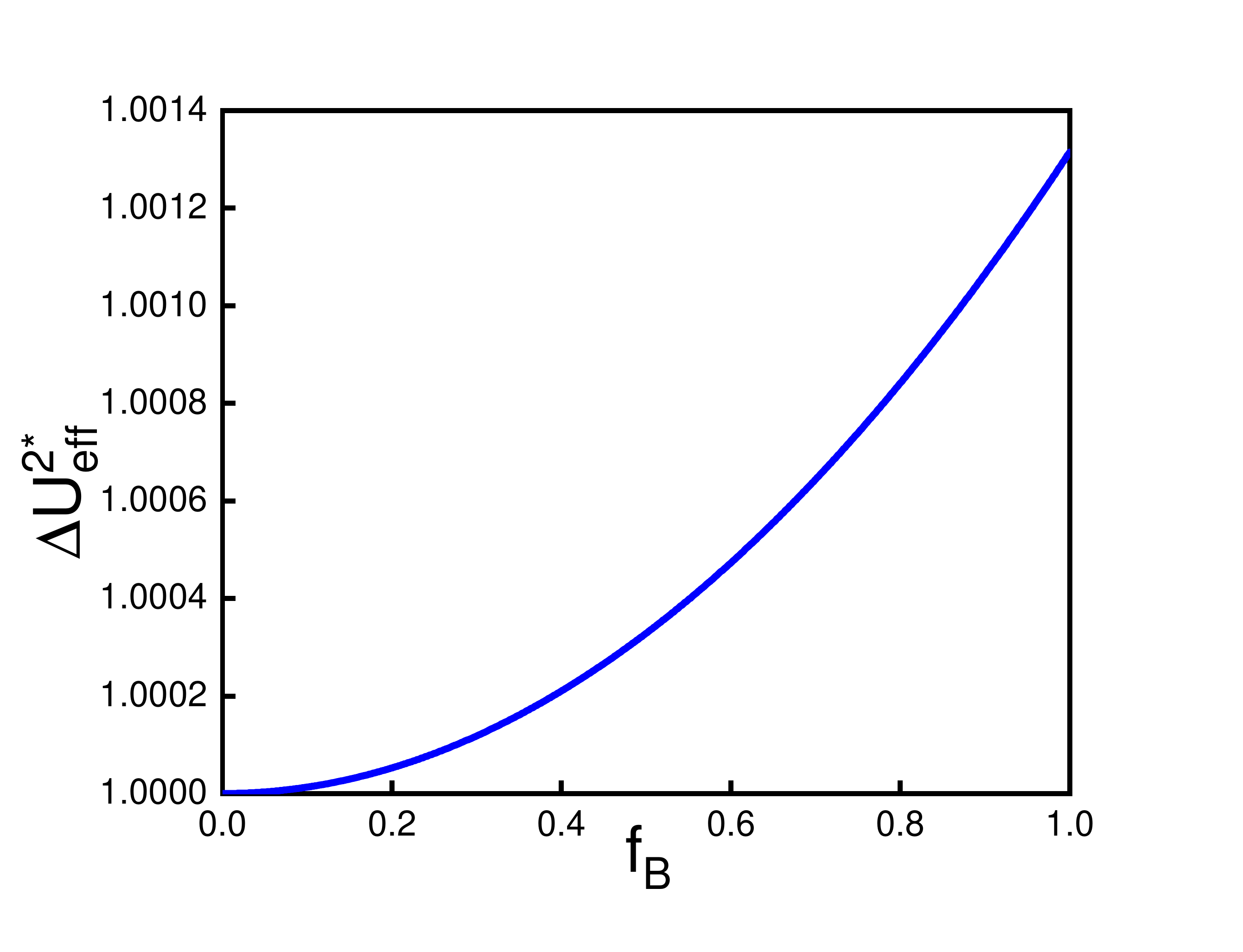} 
\end{tabular}
\end{center}
\caption{Plot of  $\Delta  U_{\textrm{eff}}^{2*}$ against $f_B$  for single particle in AREL. The values of the parameters used for both the plots are   $m=1, \gamma=1, k_B=1, T=1, a=1, b=5, \omega_{\textrm{min}}=10, \omega_{\textrm{max}}=20$.}
\label{fig:active_potential_height_B}
\end{figure*}

\noindent where

\begin{equation}
\begin{split}
\Delta  U_{\textrm{eff}}^1&= \left[-2 U_{\textrm{min}}+\frac{ U_{\textrm{min}}^2}{a^2 \left(\frac{1}{2}m  \omega_{\textrm{min}}^2-\frac{\beta  C_0}{4\tau_A}\right)}+\frac{ U_{\textrm{max}}^2}{b^2 \left(\frac{1}{2}m  \omega_{\textrm{max}}^2-\frac{\beta  C_0}{4\tau_A}\right)}\right] \\
&- \frac{1}{2\beta}\ln\left[\left(1-\frac{\beta C_0}{2 \tau_A m  \omega_{\textrm{min}}^2 }\right) \left(1-\frac{\beta C_0}{2 \tau_A m  \omega_{\textrm{max}}^2 }\right)\right]
\label{eq:active_eff_U_modelA}
\end{split}
\end{equation} 

\begin{equation}
\begin{split}
\Delta U_{\textrm{eff}}^2&= \left[-2 U_{\textrm{min}}+\frac{ U_{\textrm{min}}^2}{a^2 \left(\frac{1}{2}m  \omega_{\textrm{min}}^2-\frac{\beta  f_B^2}{4}\right)}+\frac{ U_{\textrm{max}}^2}{b^2 \left(\frac{1}{2}m  \omega_{\textrm{max}}^2-\frac{\beta f_B^2}{4}\right)}\right] \\
&- \frac{1}{2\beta}\ln\left[\left(1-\frac{\beta f_B^2}{2  m  \omega_{\textrm{min}}^2 }\right) \left(1-\frac{\beta f_B^2}{2  m  \omega_{\textrm{max}}^2 }\right)\right]
\label{eq:active_eff_U_modelB}
\end{split}
\end{equation}

\noindent  In the AREL framework, the effective barrier heights $\Delta U_{\textrm{eff}}^1$ (Eq. \ref{eq:active_eff_U_modelA}) and $\Delta U_{\textrm{eff}}^2$ (Eq. \ref{eq:active_eff_U_modelB}) depend on the active parameters $\left(C_0, f_B, \tau_A\right)$ and ambient temperature $T$ in a complex way unlike the case where the barrier height is simply shifted by a constant due to activity \cite{wexler2020dynamics}. Variation of non-dimensional effective barrier height $\left(\Delta U_{\textrm{eff}}^{1*}=\frac{\Delta U_{\textrm{eff}}^1}{\Delta U}\right)$ against $C_0$ and $\tau_A$ can be seen in Fig. \ref{fig:active_potential_height_A} for model 1 where $\Delta U=U_{\textrm{max}} - U_{\textrm{min}}$. For model 1, $\Delta U_{\textrm{eff}}^{1*}$ increases  with $C_0$ (Fig. \ref{fig:active_potential_height_A} (a)) and decreases monotonically with increasing $\tau_A$ (Fig. \ref{fig:active_potential_height_A} (b)). However for model 2, $\Delta U_{\textrm{eff}}^{2*}$ monotonically increases with  $C_0$ and independent of $\tau_B$ (Fig. \ref{fig:active_potential_height_B} ).  In the AREL framework, the effective barrier height is always greater than $\Delta U$ for both the models 1 and 2. This explains why the MFPT with non-zero activity in the AREL framework is always higher than MFPT for the zero activity situation. Here activity slows down the escape of the tagged particle.

\section{The effective diffusion coefficient for rugged energy landscape}

\noindent In ref. \cite{zwanzig1988diffusion}, Zwanzig replaced the spatial averaging of the rugged landscape by writing an effective Smoluchowski equation in steady state where the effective potential is given by
\begin{equation}
\begin{split}
 \tilde{U}(x)=U_0(x) - \frac{\lambda_-(x)}{\beta}
\label{eq:active_tau_potential}
\end{split}
\end{equation}

\noindent with effective diffusion coefficient,

\begin{equation}
\begin{split}
 \tilde{D}(x)= \frac{D}{\exp (\lambda_+ (x)) \exp (\lambda_- (x))}
\label{eq:active_tau_diffusion}
\end{split}
\end{equation}

\noindent  Eq. (\ref{eq:active_tau_final}) can be obtained by substituting Eq. (\ref{eq:active_tau_potential}, \ref{eq:active_tau_diffusion}) in Eq. (\ref{eq:active_tau}).  However, in an alternative way, one can also calculate the active MFPT for rugged energy landscape similar to Eq. (\ref{eq:active_tau_model_1}) and (\ref{eq:active_tau_model_2}) using Eq. (\ref{eq:active_tau_potential}) and (\ref{eq:active_tau_diffusion}). The readers are referred to Appendix  \ref{appendix:a} for  detailed calculations.
\\
\\
Hence when the escape time of the particle is considered in rough active potential, one can identify the effective diffusivity  for model 1,

\begin{equation}
\begin{split}
 \tilde{D}_1 (x)= D \exp \left[-\frac{\beta^2 x^2 C_0}{2\tau_A}\right]
\label{eq:active_tau_diffusion_A}
\end{split}
\end{equation}

\noindent and the same for model 2,

\begin{equation}
\begin{split}
 \tilde{D}_2 (x)= D \exp \left[-\frac{\beta^2 x^2 f_B^2}{2}\right]
\label{eq:active_tau_diffusion_B}
\end{split}
\end{equation}

\noindent In Eq. (\ref{eq:active_tau_diffusion_A}) and (\ref{eq:active_tau_diffusion_B}), the effective diffusivity, $ \tilde{D}_1(x)$ and $ \tilde{D}_2 (x)$  decreases with increasing activity ($C_0$ or $f_B$).  
However, in our AREL formalism, since the medium is dense, there is no self-propulsion velocity associated with the tagged particle, rather the activity is embedded in the dense environment, which makes the landscape rugged.

\section{Influence of additional ruggedness in active landscape}

\noindent If there is an additional rugged potential, $U_R(x)$, then we can decompose the total potential energy $(U(x))$ of the system into three separate contributions, $U(x)=U_0(x)+U_A(x)+U_R(x)$. Such models are useful to describe the dynamics of a passive polymer in active bath where the background smooth potential, $U_0(x)$, can serve the purpose of the energy bias against locally unfavorable configurations and the ruggedness comprises two distinct components: one from the difference in energies of the configurations associated with the positioning of different residues near or far from each other \cite{onuchic1997theory} which is accounted for by internal friction in polymer chains \cite{kailasham2020wet} and another from the net effect of the active forcing experienced by the particle. In this case the MFPT can be useful to describe the looping time of a polymer in an active bath \cite{shin2015facilitation}. The integral over $z$ in Eq. (\ref{eq:active_tau}) can be approximated by, 

\begin{equation}
\begin{split}
 \int dz \exp \left(-\beta U(z)\right) & \approx  \int dz \exp \left(-\beta U_0(z)\right) {\left< \exp \left(-\beta U_A(z)\right)\right>}_A  {\left< \exp \left(-\beta U_R(z)\right)\right>}_R 
\label{eq:active_tau_approx_protein}
\end{split}
\end{equation} 

\noindent where $\left<......\right>_A$ and $\left<......\right>_R$ denote the independent averages over the active noise and the small ripples of energy fluctuations due to different configurations respectively. $U_R(x)$ can be periodic function of $x$ such that $U_R(x)=\epsilon \cos(qx)$ where $\epsilon$ is  the characteristic energy scale of the potential barriers of $U_R(x)$ \cite{zwanzig1988diffusion}. When $\beta \epsilon$ is very large, then by integration over one period we obtain the $ \left<\tau (x_0)\right>$ for model 1, 

\begin{equation}
\begin{split}
 \left<\tau^1_U\right>_{\epsilon}&=\frac{\pi}{D\sqrt{\Omega_1 \Omega_2}} \exp \left[-2\beta (U_{\textrm{min}}-\epsilon)+\frac{\beta^2 U_{\textrm{min}}^2}{a^2 \Omega_1}+\frac{\beta^2 U_{\textrm{max}}^2}{b^2 \Omega_2}\right]
\label{eq:active_tau_model_1_protein}
\end{split}
\end{equation} 

\noindent Similarly for model 2, 

\begin{equation}
\begin{split}
 \left<\tau^2_U\right>_{\epsilon}&=\frac{\pi}{D\sqrt{\xi_1 \xi_2}} \exp \left(-2\beta (U_{\textrm{min}}-\epsilon)+\frac{\beta^2 U_{\textrm{min}}^2}{a^2 \xi_1}+\frac{\beta^2 U_{\textrm{max}}^2}{b^2 \xi_2}\right)
\label{eq:active_tau_model_2_protein}
\end{split}
\end{equation} 

\noindent If $U_R(x)$ is independent of $x$ and drawn form a Gaussian distribution of zero mean and variance $\sigma$ \cite{zwanzig1988diffusion}, then $\left<\tau (x_0)\right>$ for model 1, 

\begin{equation}
\begin{split}
 \left<\tau^1_U\right>_{\sigma}&=\frac{\pi}{D\sqrt{\Omega_1 \Omega_2}} \exp \left(-2\beta U_{\textrm{min}}+\frac{\beta^2 U_{\textrm{min}}^2}{a^2 \Omega_1}+\frac{\beta^2 U_{\textrm{max}}^2}{b^2 \Omega_2}+\beta^2 \sigma^2\right)
\label{eq:active_tau_model_1_Gaussian}
\end{split}
\end{equation} 

\noindent Similarly for model 2, 

\begin{equation}
\begin{split}
 \left<\tau^2_U\right>_{\sigma}&=\frac{\pi}{D\sqrt{\xi_1 \xi_2}} \exp \left(-2\beta U_{\textrm{min}}+\frac{\beta^2 U_{\textrm{min}}^2}{a^2 \xi_1}+\frac{\beta^2 U_{\textrm{max}}^2}{b^2 \xi_2}+\beta^2 \sigma^2\right)
\label{eq:active_tau_model_2_Gaussian}
\end{split}
\end{equation} 
\\
\noindent For Eq. (\ref{eq:active_tau_model_1_protein}) and (\ref{eq:active_tau_model_1_Gaussian}), the variations of $ \left<\tau^1_U\right>_{\epsilon}$  and $ \left<\tau^1_U\right>_{\sigma}$ against $C_0$ and $\tau_A$  show similar trends like the plots in Fig. \ref{fig:active_energy_A} but are shifted by a constant. Similarly for Eq. (\ref{eq:active_tau_model_2_protein}) and (\ref{eq:active_tau_model_2_Gaussian}), the variations of $ \left<\tau^2_U\right>_{\epsilon}$ and $ \left<\tau^2_U\right>_{\sigma}$ against $f_B$  show similar trends like the plot in Fig. \ref{fig:active_energy_B} but are shifted by a constant. 

\section{The role of effective temperature on escape dynamics}

\noindent For model 1, Eq. (\ref{eq:active_tau_model_1}) is valid only when both $m  \omega_{\textrm{min}}^2$ and $m  \omega_{\textrm{max}}^2$ are greater than $\frac{\beta  C_0}{2\tau_A}$ and  the same is true  for model 2 (Eq. (\ref{eq:active_tau_model_2})) with $m  \omega_{\textrm{min}}^2>\frac{\beta  f_B^2}{2}$ and $m  \omega_{\textrm{max}}^2>\frac{\beta  f_B^2}{2}$. This suggests that the activity ($C_0$ or $f_B$) should be small or  the stiffness of the potential at the minimum and maximum of the potential should be large for the  escape to  happen. However, for $\tau_A \rightarrow 0$, $\left<\tau_U\right>$ (Eq. (\ref{eq:active_tau_model_1})) is not a real quantity. To avoid this nuisance, we have to take the limit, $\tau_A \rightarrow 0$ in Eq. (\ref{eq:ornstein_corr}). Consequently, the escape will be thermally driven at  renormalized  temperature. These give us a way to characterize the MFPT using the notion of effective temperature for moderate to high activity. The active Smoluchowski equation (ASE) in the steady state is

\begin{equation}
\begin{split}
 \frac{\partial P(x,t|x_0,0)}{\partial t}&=-\frac{\partial J (x,t|x_0,0)}{\partial x}
\label{eq:active_fokker_eff}
\end{split}
\end{equation} 

\begin{equation}
\begin{split}
J  (x,t|x_0,0)=-D_{\textrm{eff}} \exp \left(- \beta_{\textrm{eff}}  U_0(x)\right) \frac{\partial }{\partial x} \left[\exp \left(\beta_{\textrm{eff}} U_0(x)\right) P(x,t|x_0,0)\right]
\label{eq:active_flux_eff}
\end{split}
\end{equation} 

\noindent where $D_{\textrm{eff}}=\frac{k_B T_{\textrm{eff}}}{\gamma}$ and $\beta_{\textrm{eff}}=\frac{1}{k_B T_{\textrm{eff}}}$. The detailed descriptions of the effective temperature and ASE for  single particle in a potential $U_0(x)$  are given in ref. \cite{chaki2019effects}. In this case, the mean escape time $\left(\left<\tau_{T_{\textrm{eff}}} \right>\right)$ over a potential barrier of height $ \Delta U_0$ immediately follows Kramers' like form \cite{kramers1990physica, chakrabarti2007exact,chaudhury2008dynamics} with an effective temperature $T_{\textrm{eff}}$, 

\begin{equation}
\begin{split}
 \left<\tau_{T_{\textrm{eff}}} \right>& = \tau_0 \exp \left(\frac{\Delta U_0}{k_B T_{\textrm{eff}}}\right) 
\label{eq:active_tau_temperature}
\end{split}
\end{equation}

\noindent where $\tau_0=\frac{2\pi \gamma}{m \omega_{\textrm{max}}\omega_{\textrm{min}}}$ and $\Delta U_0=U_{\textrm{max}}-U_{\textrm{min}}$. Here we assume that there will be an effective equilibrium in the harmonic well. For model 1, $T_{\textrm{eff}}$ is denoted by $T^1_{\textrm{eff}}$ and thus, $T^1_{\textrm{eff}}=T+\frac{ C_0}{k_B \gamma \tau_A \left(\frac{m \omega_{\textrm{min}}^2}{\gamma}+\frac{1}{\tau_A}\right)}$. Similarly for model 2, $T^2_{\textrm{eff}}=T+\frac{ f_B^2}{\gamma k_B \left(\frac{m \omega_{\textrm{min}}^2}{\gamma}+\frac{1}{\tau_B}\right)}$. For model 1 and 2, $\left<\tau_{T_{\textrm{eff}}} \right>$ are denoted by  $\left<\tau^1_{T_{\textrm{eff}}} \right>$ and  $\left<\tau^2_{T_{\textrm{eff}}} \right>$ respectively. 

\begin{figure*}[ht]
\begin{center}
\begin{tabular}{cc}
\includegraphics[width=0.55\textwidth]{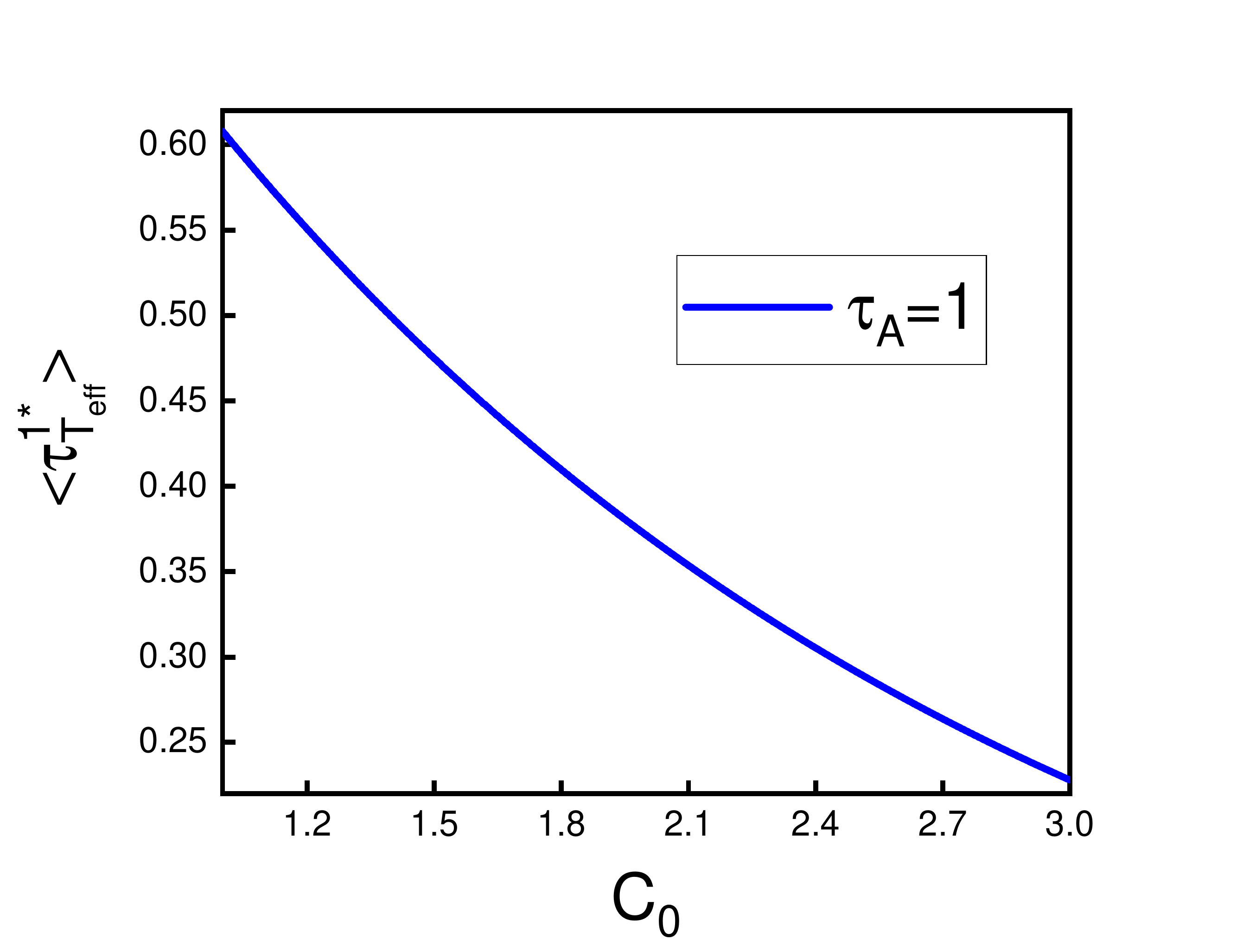} &
\includegraphics[width=0.55\textwidth]{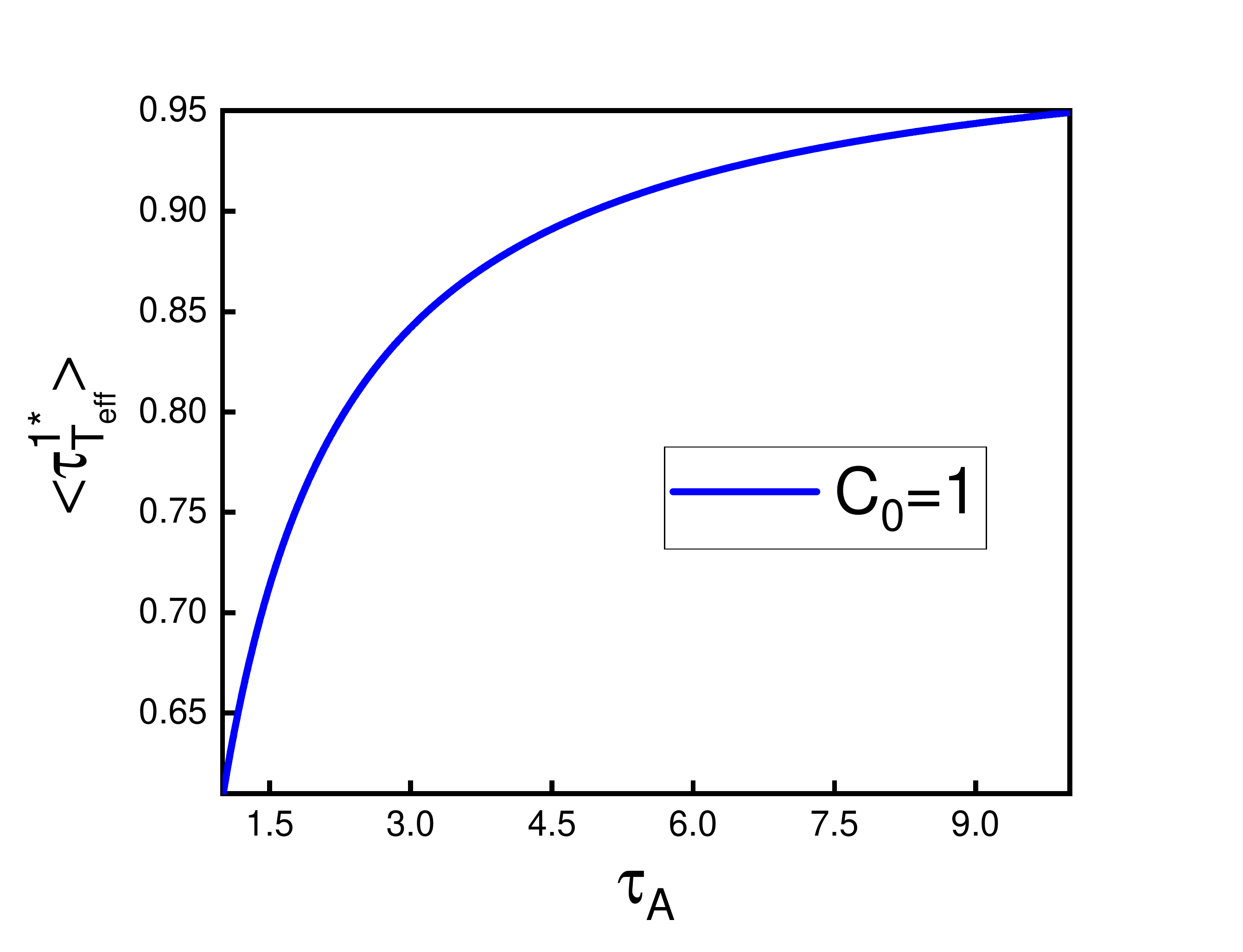}\\
(a) & (b)  \\
\end{tabular}
\end{center}
\caption{Plots of $\left<\tau^{1*}_{T_{\textrm{eff}}} \right>$ against $C_0$ (plot (a)) and $\tau_A$ (plot (b)) for single particle in AREL. The values of the parameters used for both the plots are   $m=1, \gamma=1, k_B=1, T=10, a=1, b=5, \omega_{\textrm{min}}=20, \omega_{\textrm{max}}=10$.}
\label{fig:active_eff_temp_A}
\end{figure*}

\begin{figure*}[ht]
\begin{center}
\begin{tabular}{cc}
\includegraphics[width=0.55\textwidth]{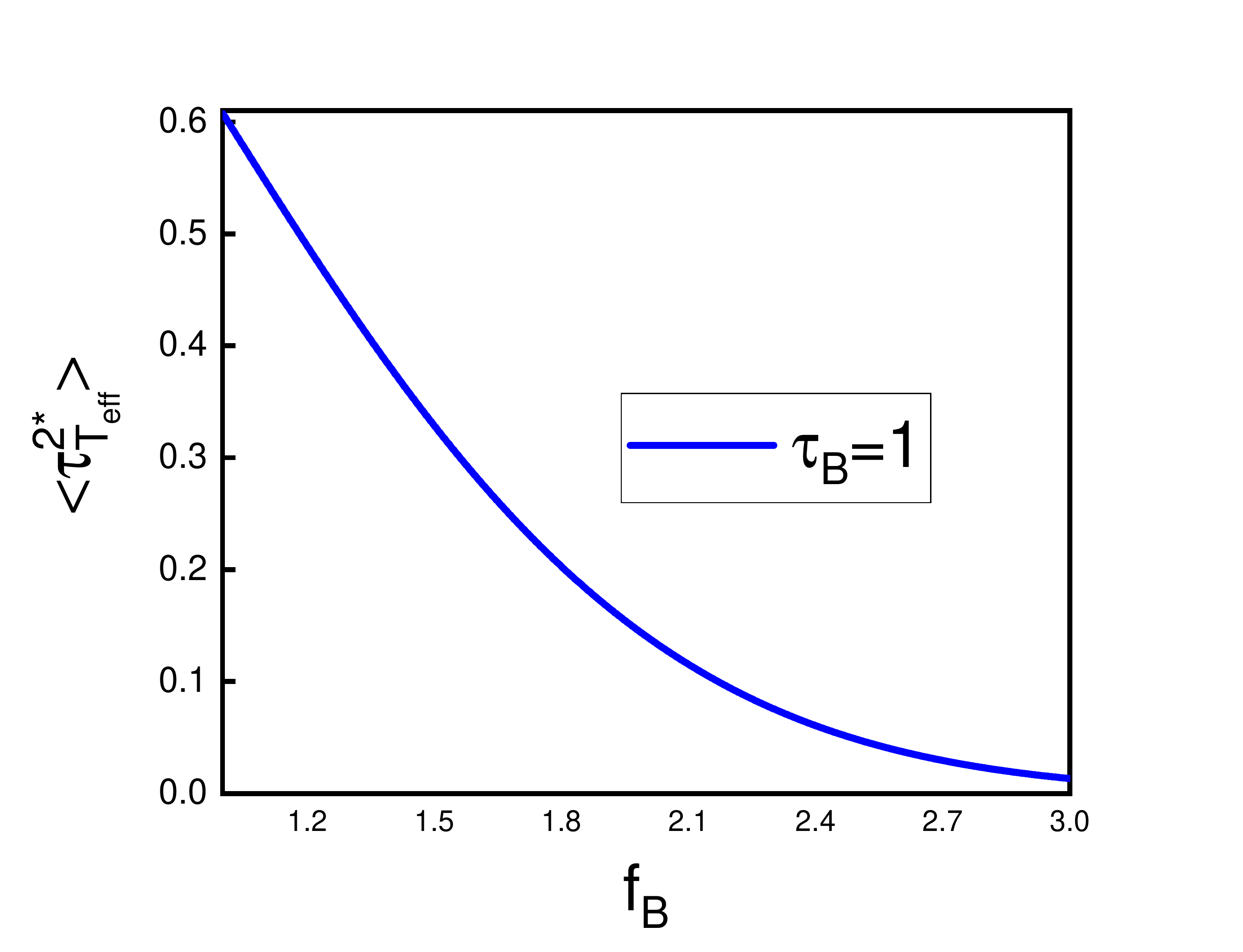} &
\includegraphics[width=0.55\textwidth]{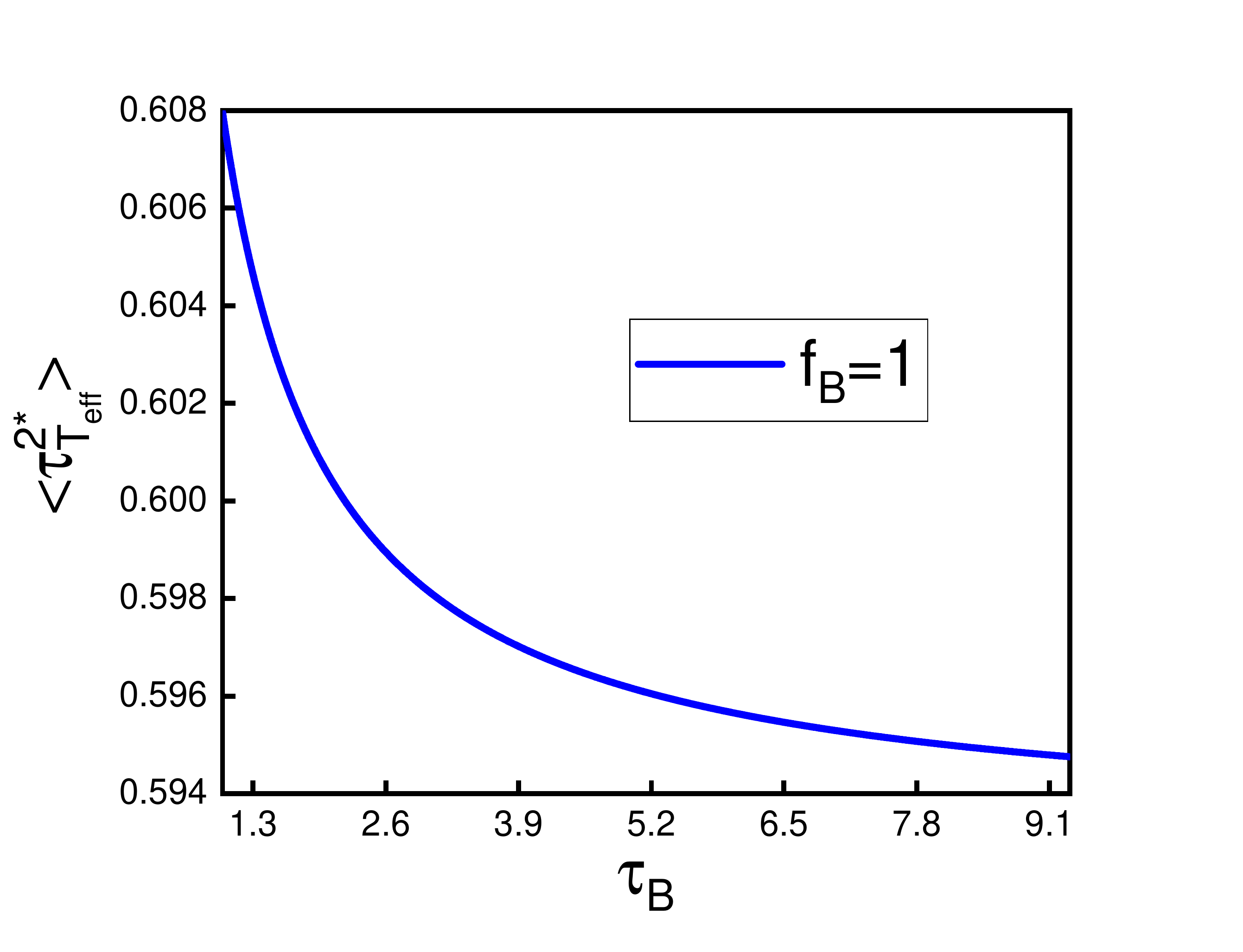}\\
(a) & (b)  \\
\end{tabular}
\end{center}
\caption{Plots of $\left<\tau^{2*}_{T_{\textrm{eff}}} \right>$ against $f_B$ (plot (a)) and $\tau_B$ (plot (b)) for single particle in AREL. The values of the parameters used for both the plots are   $m=1, \gamma=1, k_B=1, T=10, a=1, b=5, \omega_{\textrm{min}}=20, \omega_{\textrm{max}}=10$.}
\label{fig:active_eff_temp_B}
\end{figure*}

\noindent Variations of non-dimensional MFPTs $\left(\left<\tau^{1*}_{T_{\textrm{eff}}} \right>=\frac{\left<\tau^1_{T_{\textrm{eff}}} \right>}{\left<\tau_T \right>}\right)$ against $C_0$ and $\tau_A$ and $\left(\left<\tau^{2*}_{T_{\textrm{eff}}} \right>=\frac{\left<\tau^2_{T_{\textrm{eff}}} \right>}{\left<\tau_T \right>}\right)$ against $f_B$ can be seen in Fig. \ref{fig:active_eff_temp_A}  and \ref{fig:active_eff_temp_B} respectively. For both the models 1 and 2, with increasing the activity ($C_0$ or $f_B$), $T_{\textrm{eff}}$ increases, resulting in a fast escape of the tagged particle ( see Fig. \ref{fig:active_eff_temp_A} (a) and \ref{fig:active_eff_temp_B} (a)). However,  increasing $\tau_A$ for model 1 would have the same effect as decreasing $\tau_B$ for model 2 on the MFPTs ( see Fig. \ref{fig:active_eff_temp_A} (b) and \ref{fig:active_eff_temp_B} (b)). A similar trend in the noise strength and persistence time on the dynamics of active glass has been observed by Nandi $et.\, al.$ \cite{nandi2018random}.

\section{Physical origin of the AREL framework and the effective temperature description}

\noindent In the following, we will discuss the range of validity of the AREL framework and the effective temperature description to calculate MFPTs. In the absence of any non-equilibrium activity, the tagged particle moves in a smooth potential which has a harmonic trap and an inverted harmonic barrier at a point far from the minima of the trap \cite{chakrabarti2007exact}. This  potential depends on the mean inter-particle interactions, density etc. However, the inclusion of a small but non-zero activity perturbs the local arrangements of the surrounding particles and the rate of rearrangement is very slow in a dense environment. Hence, the tagged particle diffuses in a rugged energy landscape (Fig. \ref{fig:schematic} (a)) and this ruggedness caused by the activity in steady state, slows down the dynamics. However, the escape from a rugged energy landscape ( Eq. (\ref{eq:active_tau_model_1}, \ref{eq:active_tau_model_2})) is valid only for small activity. The upper bounds of the activity for model 1, $C_0 \leq 2 k_B T \tau_A m \omega^2_{\textrm{max}}$ and $C_0 \leq 2 k_B T \tau_A m \omega^2_{\textrm{min}}$. The same for model 2, $f_B \leq \sqrt{2 k_B T m \omega^2_{\textrm{max}}}$ and $f_B \leq \sqrt{2 k_B T m \omega^2_{\textrm{min}}}$. Hence, the active rugged energy landscape description seems consistent in the limit of small activity. For model 1, the local environment of the tagged particle becomes less disordered with increasing persistence time $(\tau_A)$ \cite{szamel2015glassy, flenner2016nonequilibrium}. Hence, MFPT for model 1 decreases as the persistence time increases. In contrast, with increasing activity, the surrounding active particles move faster than the passive particles. Hence, the surroundings of the tagged particle becomes a dilute active medium and the structural rearrangements due to activity is minimal. However, we assume that the tagged particle will experience a harmonic trap of finite depth due to an averaged many-body interactions (Fig. \ref{fig:schematic} (b)) and such an effective harmonic potential is used in many different contexts to describe the tagged particle dynamics, such as in cell membrane \cite{gov2003cytoskeleton}, elastic gel network \cite{ben2015modeling} or in active glass \cite{nandi2018random, nandi2017nonequilibrium, mandal2016active,shen2004stability}. In this case, the notion of effective temperature is quite useful to describe the fluctuations coming from the medium and as a result the mean escape time from the trap will be short at relatively higher activity as evident from Eq. (\ref{eq:active_tau_temperature}).  
\\
\\
Recently, Leocmach $et. al.$ have experimentally showed a non-monotonic behavior of relaxation time with activity for glassy dynamics: it increases first for small but non-zero activity, then decreases with high activity \cite{klongvessa2019active}. The relaxation time in dense systems is the reminiscent of MFPT over a potential barrier. For high activity limit, the relaxation time decreases which signifies enhanced motion of the particles. Hence, at high activity limit, the MFPT can be well described by the effective temperature approach. However, the dense passive system relaxes due to isotropic cooperative motion which is well formulated by incorporating the concept of smooth elastic (or harmonic) potential barrier as suggested by Mirigan and Schweizer \cite{mirigian2014elastically}. In the presence of small but non-zero activity, Leocmach $et. al.$ have observed slowdown to a drop in efficiency of cooperative relaxations. Here, we propose that  the smooth landscape becomes rugged due to small activity and consequently, in the low activity limit, the dynamics becomes slow with increasing the activity as shown in Fig. \ref{fig:active_energy_A} (a) and Fig. \ref{fig:active_energy_B}. In Ref. \cite{klongvessa2019active}, increase in the relaxation time up to a factor of $10$ has been observed for a range of activities, when the density is high and the activity is small. Similarly, in our AREL approach we see an increase of the MFPT. Depending on the choice of parameters, this increase is by a factor of $1.2$ ($m=1, \gamma=1, k_B=1, T=1, a=1, b=5, \omega_{\textrm{min}}=10, \omega_{\textrm{max}}=20, \tau_A=10$) or $10$ ($m=1, \gamma=1, k_B=1, T=1, a=10, b=15, \omega_{\textrm{min}}=10, \omega_{\textrm{max}}=10, \tau_A=15$). 
\\
\\
In a dense system with low activity, the slowing down of the tagged particle dynamics with activity is different from the jamming behavior in dense active systems with small activity and high persistence time, as observed in simulations \cite{mandal2020extreme}.  This is the case of extreme active matter. In our AREL formalism, activity ($C_0$) induces ruggedness and hence the dynamics slows down. On the other hand, the system becomes less jammed, due to enhanced motion of the tagged particle on increasing the activity, as shown by Mandal $et. al$ \cite{mandal2020extreme}. In a future work, we would like to extend our formalism to extreme active matter.

\section{Discussion and Conclusions}

\noindent The timescale with which a Brownian particle relaxes to equilibrium in a harmonic trap with stiffness $k$ is $\frac{\gamma}{k}$, where $\gamma$ is the friction experienced by the Brownian particle. If the potential is modeled as an inverted harmonic one, at a point far from the minima of the trap, then the mean time to reach the top of this inverted harmonic potential is much higher than the equilibration time scale $\frac{\gamma}{k}$, as it involves climbing up a barrier of several $k_B T$. In such a situation, a well defined rate of escape exists, which can be expressed either as the mean first passage time or by a steady state flux over population description. Though this time scale separation is present in a steady state of active systems, the mean escape time \cite{wexler2020dynamics, caprini2019active} significantly deviates from the well known Kramers' rate, thus reflecting the breakdown of detailed balance in active systems. We have mapped the Zwanzig's theory of MFPT based on a rugged energy landscape to a mixture of active and passive particles where the density of active particles is very small compared to the density of passive particles. In our model system, the ruggedness is induced by the activity and the smooth background trapping potential comes from the interaction between the passive particles. Like the kinetic model developed by Chakrabarti and Bagchi in the context of glassy dynamics, where the $\alpha$-relaxation is described as a concerted series of $\beta$-relaxation mediated cooperative transitions in a double well \cite{chakrabarti2005frequency}, our AREL model accounts for the escape of a tagged passive particle on a rugged energy landscape where the ruggedness comes from activity. The AREL approach is only valid on the time scale shorter than the persistence time. When activity evolves on the time scale of persistence time, the system shows directed motion instead to diffusion. We put forward an analytic expression of the MFPT for AREL which is higher than the  Kramers' theory of MFPT. This is a signature of the inherent non-equilibrium nature of the active matter. We consider two different models of active noise statistics: model 1 and 2. However, for both the models 1 and 2, the activity ($C_0$ or $f_B$) facilitates the ruggedness and thus, the MFPT increases with increasing the activity. This behavior is fundamentally different from the studies \cite{wexler2020dynamics, caprini2019active} on the long time dynamics of active systems where the dynamics is always faster due to activity. In both the models 1 and 2, the temporal correlations of the active noise decay exponentially. But, the strength of the active noise has different interpretations for models 1 and 2. However, the precise form of the correlation function of the active noise will strongly affect the MFPT  in AREL.  For model 1,  the MFPT decreases with increasing the persistence time $\tau_A$. This is because of the fact that longer persistence time makes  the surroundings of the tagged particle more ordered which essentially diminishes the ruggedness. But for  model 2, the MFPT is independent of $\tau_B$. This demonstrates that the result will depend on the microscopic details of the activity. From the expression of MFPT, we have defined an effective barrier height, $\Delta U^1_\textrm{eff}$ for model I  and $\Delta U^2_\textrm{eff}$ for model II. With activity, $\Delta U^1_\textrm{eff}$ and $\Delta U^2_\textrm{eff}$ are higher compared to the barrier height $\Delta U$ in standard Kramers' expression and hence the MFPT calculated from the AREL framework is always greater than Kramers' MFPT. 
\\
\\
\noindent In the context of polymers, there will be an additional ruggedness associated with the various interactions due to different conformational changes and hence the dynamics will be further slowed down. A natural question arises as to what the effective diffusion coefficient for the slow dynamics will be. We have proposed an effective diffusion coefficient for the dynamics in AREL. For both the models, $\tilde{D} (x)$ decreases with increasing the activity which makes it appropriate to analyze the activity-induced slow dynamics.  However for high temperature, the MFPTs for both models 1 and 2 become Kramers' MFPT though the active particles are still present in the medium. On the other hand, for high activity, the expression for the MFPT becomes imaginary.  Thus, the AREL approach to describe the slow dynamics is valid for small activity and low temperature (or high stiffness) limit. Ray $et. \, al.$ have analyzed the  Kramers' escape rate due to an external non-equilibrium load \cite{mondal2009kramers}. The load controls the active  transport of biological motor proteins \cite{mondal2009kramers, fodor2015activity, fodor2014energetics}. In their model, the load only affects the pre-factor of the Kramers' mean escape rate. However, in our AREL approach, activity modifies the pre-factor as well as the exponential dependence of the barrier height of Kramers' MFPT. For high activity, the active particles move faster than the passive particles. Hence, the medium becomes a dilute active medium but the background caging potential created by the surrounding passive particles  will be unaffected at least qualitatively, however, the stiffness will change as compared to a dense system. A common approach is to define an effective temperature derived from the equipartition theorem for the position of the particle in steady state and use it to calculate the MFPT. For both the models 1 and 2, the activity plays the same role: the MFPT decreases with increasing the activity. However, the behavior of MFPT with respect to $\tau_A$ in model 1 is quite opposite to that of the $\tau_B$ in model 2. For a dense medium with low activity, it will be increasingly difficult for the tagged particle to move. For such systems, the applicability of effective temperature has been shown to be limited \cite{lowen2020inertial,fily2012athermal}.
\\
\\
The self-diffusivity, $D_{\textrm{self}}$, in a rugged energy landscape is well connected with the excess entropy, $S_{\textrm{ex}}$, in terms of an exponential relation, $D_{\textrm{self}} \sim e^{\Lambda S_{\textrm{ex}}}$ \cite{seki2015relationship}. Computing excess entropy from our AREL approach will be quite useful in the context of non-equilibrium active liquids to study their dependence on activity.  The present form of AREL framework describes the static disorder in the energy landscape due to activity. It will be interesting to extend the AREL approach to the more generic case of dynamical disorder \cite{Acharya2017, Kwon2014, chechkin2017brownian, Jain2016, tyagi2017non, zwanzig1990rate, debnath2006rate}.

\section{Acknowledgements}

\noindent We thank R. Kailasham for critical reading of the manuscript. We would also extend our gratitude to the anonymous reviewers for their useful comments. SC thanks DST Inspire for the fellowship. RC acknowledges SERB (Project No. SB/SI/PC-55/2013) and IRCC-IIT Bombay (Project No. RD/0518-IRCCAW0-001) for funding.

\appendix

\section{ ALTERNATIVE DERIVATION OF MFPT FOR RUGGED ENERGY LANDSCAPE} 

\noindent The effective Smoluchowski equation is given by,

\begin{equation}
\begin{split}
 \frac{\partial P(x,t|x_0,0)}{\partial t}&=-\frac{\partial J (x,t|x_0,0)}{\partial x}
\label{eq:active_fokker_cal}
\end{split}
\end{equation} 

\begin{equation}
\begin{split}
J  (x,t|x_0,0)&=-\tilde{D} \exp \left(-\beta \tilde{U}(x)\right) \frac{\partial }{\partial x} \left[\exp \left(\beta \tilde{U}(x)\right) P(x,t|x_0,0)\right]\\
&=-D \exp (-\lambda_+ (x)) \exp (-\lambda_- (x)) \exp \left(-\beta \left(U_0(x) - \frac{\lambda_-(x)}{\beta}\right)\right)\\
& \times \frac{\partial }{\partial x} \left[\exp \left(\beta \left(U_0(x) - \frac{\lambda_-(x)}{\beta}\right)\right) P(x,t|x_0,0)\right] \\
&=-D \exp \left(-\lambda_+ (x)) -\beta U_0(x) \right) \frac{\partial }{\partial x} \left[\exp \left(\beta U_0(x) - \lambda_-(x)\right) P(x,t|x_0,0)\right]
\label{eq:active_flux_cal}
\end{split}
\end{equation} 

\noindent In steady state, $\frac{\partial P(x,t|x_0,0)}{\partial t} \approx 0$ and hence, $J$ will be independent of $x$. From Eq. (\ref{eq:active_flux_cal}) we get,

\begin{equation}
\begin{split}
&-\frac{J \exp \left(\lambda_+ (x)) +\beta U_0(x) \right)}{D}= \frac{\partial }{\partial x} \left[\exp \left(\beta U_0(x) - \lambda_-(x)\right) P(x)\right] \\
&-\frac{J}{D}\int_a^c dx \exp \left(\lambda_+ (x)) +\beta U_0(x) \right) =\left[\exp \left(\beta U_0(x) - \lambda_-(x)\right) P(x)\right]_a^c
\label{eq:active_flux_cal_1}
\end{split}
\end{equation} 

\noindent $P(x)$ is very small at $c$. Hence, 

\begin{equation}
\begin{split}
& \frac{J}{D}\int_a^c dx \exp \left(\lambda_+ (x)) +\beta U_0(x) \right) =\exp \left(\beta U_0(a) - \lambda_-(a)\right) P(a) \\
&J=\frac{\exp \left(\beta U_0(a) - \lambda_-(a)\right) P(a) D}{\int_a^c dx \exp \left(\lambda_+ (x)) +\beta U_0(x) \right) }
\label{eq:active_flux_cal_2}
\end{split}
\end{equation} 

\noindent If the barrier is high, then around $a$, the current is almost zero. This defines an effective equilibrium condition in the neighborhood of $a$. In other words, this asks for a local equilibriation before the particle escapes over, even in the presence of activity. Thus, 

\begin{equation}
\begin{split}
& \frac{\partial }{\partial x} \left[\exp \left(\beta \tilde{U}(x)\right) P(x)\right]=0 \\
& \exp \left(\beta \tilde{U} (x)\right) \frac{\partial P(x)}{\partial x}+ \beta \exp \left(\beta \tilde{U} (x)\right) \frac{\partial \tilde{U} (x)}{\partial x} P(x)=0\\
&\frac{\partial P(x)}{\partial x}=- \beta \frac{\partial \tilde{U} (x)}{\partial x} P(x)   
\label{eq:active_flux_cal_3}
\end{split}
\end{equation} 

\noindent Integrating between $a$ to $x$,
\begin{equation}
\begin{split}
P(x)=P(a) \exp \left[\beta \left(\tilde{U} (a)- \tilde{U} (x)\right)\right]
\label{eq:active_flux_cal_4}
\end{split}
\end{equation} 

\noindent The  population $(n)$ in the  well is given by

\begin{equation}
\begin{split}
n=\int_{a-\Delta}^{a+\Delta} dx P(x)&=P(a) \exp \left[\beta \tilde{U} (a) \right] \int_{a-\Delta}^{a+\Delta} dx \left[-\beta \tilde{U} (x) \right]\\
&=P(a) \exp \left[\exp\left(\beta U_0(a) - \lambda_-(a)\right) \right] \int_{a-\Delta}^{a+\Delta} dx \left[\exp \left(-\beta U_0(x) + \lambda_-(x)\right) \right]\\
\label{eq:active_flux_cal_5}
\end{split}
\end{equation} 

\noindent The rate of escape $\kappa$ is thus given by,

\begin{equation}
\begin{split}
\kappa&=\frac{J}{n}\\
&=\frac{D}{\int_a^c dx \exp \left(\lambda_+ (x)) +\beta U_0(x) \right) \int_{a-\Delta}^{a+\Delta} dx \left[\exp \left(-\beta U_0(x) + \lambda_-(x)\right) \right]}
\label{eq:active_flux_cal_6}
\end{split}
\end{equation}

\noindent Hence the MFPT,
\begin{equation}
\begin{split}
\frac{1}{\kappa}= \left<\tau \right>&= \frac{1}{D} \int_a^c dy \exp \left(\lambda_+ (y)) +\beta U_0(y) \right) \int_{a-\Delta}^{a+\Delta} dz \left[\exp \left(-\beta U_0(z) + \lambda_-(z)\right) \right]
\label{eq:active_tau_cal}
\end{split}
\end{equation}

\noindent  When $k_B T$ is small (or the barrier is high), the predominant contribution to the integration over $z$ in Eq.  (\ref{eq:active_tau_cal}) comes from the immediate neighborhood of $a$ where $a$ is a simple minimum of $U(z)$,  

\begin{equation}
U_0(x)=U_{\textrm{min}}+\frac{1}{2}m \omega_{\textrm{min}}^2 \left(z-a\right)^2 +......
\label{eq:potential_z_cal}
\end{equation}

\noindent where $U_{\textrm{min}}=\frac{1}{2}m \omega_{\textrm{min}}^2 a^2$. Next we extend both the limits of the integration from $-\infty$ to $\infty$ and the integral for model 1 is, 

\begin{equation}
\begin{split}
& \int_{-\infty}^{\infty} dz \exp \left(-\beta\left( U_{\textrm{min}}+\frac{1}{2}m \omega_{\textrm{min}}^2 \left(z-a\right)^2\right)+\frac{\beta^2 z^2 C_0}{4\tau_A}\right)  \\
&=\exp \left(-2\beta U_{\textrm{min}}+\frac{\beta^2 U_{\textrm{min}}^2}{a^2 \Omega_1}\right)  \sqrt{\frac{\pi}{\Omega_1}} \\ 
\label{eq:potential_z_integration_cal}
\end{split}
\end{equation}

\noindent where $\Omega_1=\left(\frac{1}{2}m \beta \omega_{\textrm{min}}^2-\frac{\beta^2  C_0}{4\tau_A}\right)$. The absorbing barrier is placed at the maximum, $x=b$ of the potential $U(x)$. The integral over $y$ is dominated by the potential near the barrier and we will follow the same the expansion will be quadratic, 

\begin{equation}
U_0(y)=U_{\textrm{max}}-\frac{1}{2}m \omega_{\textrm{max}}^2 \left(y-b\right)^2 
\label{eq:potential_y_cal}
\end{equation}

\noindent where $U_{\textrm{max}}=\frac{1}{2}m \omega_{\textrm{max}}^2 b^2$ and set the range of integration from $-\infty$ to $\infty$. Thus, 

\begin{equation}
\begin{split}
& \int_{-\infty}^{\infty} dy \exp \left(\beta\left( U_{\textrm{max}}-\frac{1}{2}m \omega_{\textrm{max}}^2 \left(y-b\right)^2\right)+\frac{\beta^2 y^2 C_0}{4\tau_A}\right)  \\
&=\exp \left(\frac{\beta^2 U_{\textrm{max}}^2}{b^2 \Omega_2}\right)  \sqrt{\frac{\pi}{\Omega_2}} \\
\label{eq:potential_y_integration_cal}
\end{split}
\end{equation}

\noindent where $\Omega_2=\left(\frac{1}{2}m \beta \omega_{\textrm{max}}^2-\frac{\beta^2  C_0}{4\tau_A}\right)$. Substituting Eq. (\ref{eq:potential_z_integration_cal}, \ref{eq:potential_y_integration_cal}) into Eq. (\ref{eq:active_tau_cal}), we obtain

\begin{equation}
\begin{split}
 \left<\tau^1_U\right>&=\frac{\pi}{D\sqrt{\Omega_1 \Omega_2}} \exp \left(-2\beta U_{\textrm{min}}+\frac{\beta^2 U_{\textrm{min}}^2}{a^2 \Omega_1}+\frac{\beta^2 U_{\textrm{max}}^2}{b^2 \Omega_2}\right)
\label{eq:active_tau_model_1_cal}
\end{split}
\end{equation} 
\\
\\
\noindent In a similar way, one can calculate the MFPT for model 2, 

\begin{equation}
\begin{split}
 \left<\tau^2_U \right>&=\frac{\pi}{D\sqrt{\xi_1 \xi_2}} \exp \left(-2\beta U_{\textrm{min}}+\frac{\beta^2 U_{\textrm{min}}^2}{a^2 \xi_1}+\frac{\beta^2 U_{\textrm{max}}^2}{b^2 \xi_2}\right)
\label{eq:active_tau_model_2_cal}
\end{split}
\end{equation} 
\label{appendix:a}

\end{document}